%% file: ALCHEMI_HOCOp.tex
\documentclass[twocolumn]{aastex631}
\usepackage{comment,amsmath}

\newcommand{\hocop}{HOCO$^+$}
\newcommand{\cotw}{CO$_2$}
\newcommand{\ps}{s$^{-1}$}
\newcommand{\methanol}{CH$_3$OH}
\newcommand{\water}{H$_2$O}
\newcommand{\httcop}{H$^{13}$CO$^+$}
\newcommand{\htw}{H$_2$}
\newcommand{\enum}[1]{$10^{#1}$}
\newcommand{\fnum}[2]{$#1\times 10^{#2}$}
\newcommand{\kms}{km\,s$^{-1}$}
\shorttitle{Protonated \cotw\ in NGC~253}
\shortauthors{Harada et al.}

\begin{document}

\title{ALCHEMI finds a ``shocking" carbon footprint in the starburst galaxy NGC~253}

\input{ALCHEMICollabAuthList}



\begin{abstract}
Centers of starburst galaxies may be characterized by a specific gas and ice chemistry due to their gas dynamics and the presence of various ice desorption mechanisms. This may result in a peculiar observable composition. We analyze abundances of \cotw, a reliable tracer of ice chemistry, from data collected as part of the ALMA large program ALCHEMI, a wide-frequency spectral scan toward the starburst galaxy NGC~253 with an angular resolution of 1.6$\arcsec$. We constrain the \cotw\ abundances in the gas phase using its protonated form \hocop. The distribution of \hocop\ is similar to that of methanol, which suggests that \hocop\ is indeed produced from the protonation of \cotw\ sublimated from ice. The \hocop\ fractional abundances are found to be \fnum{(1-2)}{-9} at the outer part of the central molecular zone (CMZ), while they are lower ($\sim$\enum{-10}) near the kinematic center. This peak fractional abundance at the outer CMZ is comparable to that in the Milky Way CMZ, and orders of magnitude higher than that in Galactic disk star-forming regions. From the range of \hocop/\cotw\ ratios suggested from chemical models, the gas-phase \cotw\ fractional abundance is estimated to be \fnum{(1-20)}{-7} at the outer CMZ, and orders of magnitude lower near the center. We estimate the \cotw\ ice fractional abundances at the outer CMZ to be \fnum{(2-5)}{-6} from the literature. A comparison between the ice and gas \cotw\ abundances suggests an efficient sublimation mechanism. This sublimation is attributed to large-scale shocks at the orbital intersections of the bar and CMZ.
\end{abstract}



\section{Introduction} \label{sec:intro}

The abundance of interstellar molecules depends on the balance between their formation and destruction processes in the gas phase and on grain surfaces. Exchange processes can transform molecules in one phase to the other; gas-phase molecules can freeze onto dust grains as ice (adsorption), while molecules on grain surfaces can sublimate into the gas phase (desorption). Knowing both gas- and ice-phase abundances is, therefore, necessary for a comprehensive understanding of the chemical composition and their related physical conditions. Carbon dioxide (\cotw) is one of the most dominant forms of ice on interstellar dust \citep{2011ApJ...740..109O} together with H$_2$O and CO. \cotw\ is, in fact, one of the ice species detectable in extragalactic sources \citep[e.g., ][]{2007ApJ...659..296L, 2015ApJ...807...29Y}. It is obvious from the strong CO rotational emission that the gas-phase CO is abundant, but \cotw\ is thought to reside more preferentially on dust due to the inefficient gas-phase formation \citep{2011ApJ...735...15G}. Because \cotw\ has higher desorption energy than that of CO, the presence of abundant gas-phase \cotw\ requires stronger desorption mechanisms. While \cotw\ can be detected via rotational-vibrational transitions in warm gas ($\gtrsim$ several 100 K) \citep[e.g., ][]{2003Boonman} or in ice with broader line features \citep[e.g., ][]{2015ApJ...807...29Y}, it cannot be observed in cold gas due to the lack of a permanent dipole moment. 

Although the gas-phase abundances of such species without a dipole moment cannot be directly measured through commonly observed rotational transitions, it has been proposed that they can be estimated from their protonated forms \citep{1977ApJ...215..503H,2015A&A...579L..10A,2019MNRAS.483L.114R}. The protonated form of \cotw, \hocop, was first detected in Sgr B2 by \citet{1981ApJ...246L..41T}, but its line identification required spectroscopic confirmation by \citet{1982ApJ...254..405D}. Since then, \hocop\ has been detected in translucent clouds \citep{turner_physics_1999}, low-/high-mass star-forming regions \citep{sakai_detection_2008,vastel_abundance_2016,majumdar_detection_2018,fontani_protonated_2018}, the Galactic Center \citep[Sgr A and B2 clouds;][]{minh_observations_1988,minh_abundance_1991}, starburst galaxies \citep{2006ApJS..164..450M,2015AA...579A.101A} and a $z\sim0.9$ molecular absorber \citep{2013A&A...551A.109M}. Among them, \citet{minh_abundance_1991} and \citet{2015MNRAS.446.3842A} found about 2 orders of magnitude higher fractional abundances of \hocop\ in the Galactic Center than in spiral-arm molecular clouds. 
    
The main formation paths of \hocop\ are gas-phase reactions: protonation of \cotw\ such as
\begin{equation}
    {\rm CO_2 + H_3^+ \longrightarrow HOCO^+ + H_2,} \nonumber
\end{equation}
or an ion-neutral reaction
\begin{equation}
    {\rm HCO^+ + OH \longrightarrow HOCO^+ + H} \nonumber
\end{equation} 
\citep{vastel_abundance_2016,2017AandA...602A..34B}. In the former route, the \hocop\ abundance can increase due to the evaporation of \cotw\ from grain surfaces because \cotw\ is one of the most abundant forms of carbon on grain surfaces. The ice can sublimate thermally (e.g., in the vicinity of protostars), or non-thermally (e.g., photodesorption, cosmic-ray-induced evaporation, or shock sputtering). While the latter route from HCO$^+$ is considered dominant in at least some parts of high- or low-mass star-forming regions \citep{majumdar_detection_2018,fontani_protonated_2018}, the protonation of \cotw\ can be the dominant route when there is a fast mechanism of \cotw\ sublimation.

Shocks, one of drivers of the ice sublimation process, are ubiquitous in galactic centers. In many barred-spiral galaxies, galactic centers host intersections of $x_1$ and $x_2$ orbits\footnote{Bars lie on the $x_1$ orbits, while $x_2$ orbits form inner nuclear rings. The location of nuclear rings may correspond to that of inner Lindblad resonances, but this is not always the case \citep{2012ApJ...758...14K}.} \citep{1992MNRAS.259..345A,2013ApJ...769..100S,2020MNRAS.494.6030S}.
At these orbital intersections, shocks are naturally expected. Abundances of typical shock molecular tracers such as CH$_3$OH and SiO have been found to be enhanced at locations of orbital intersections in IC342 and M83 \citep{2005ApJ...618..259M,2019ApJ...884..100H}.

\begin{figure*}[h]
\centering{
\includegraphics[width=0.48\textwidth]{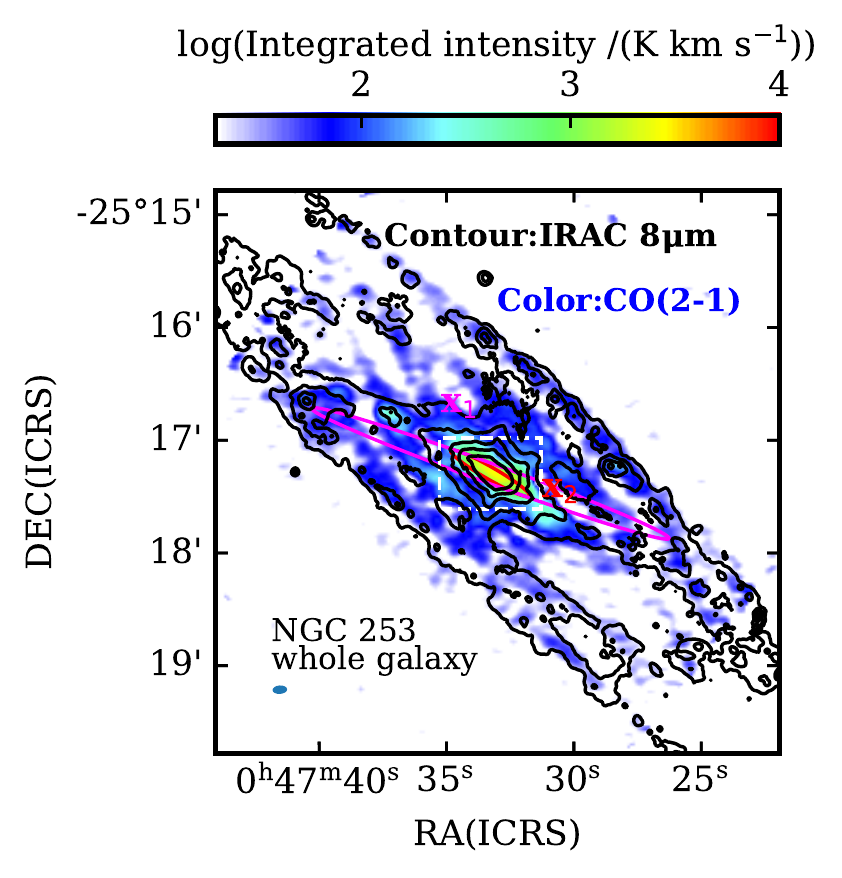}
\includegraphics[width=0.48\textwidth]{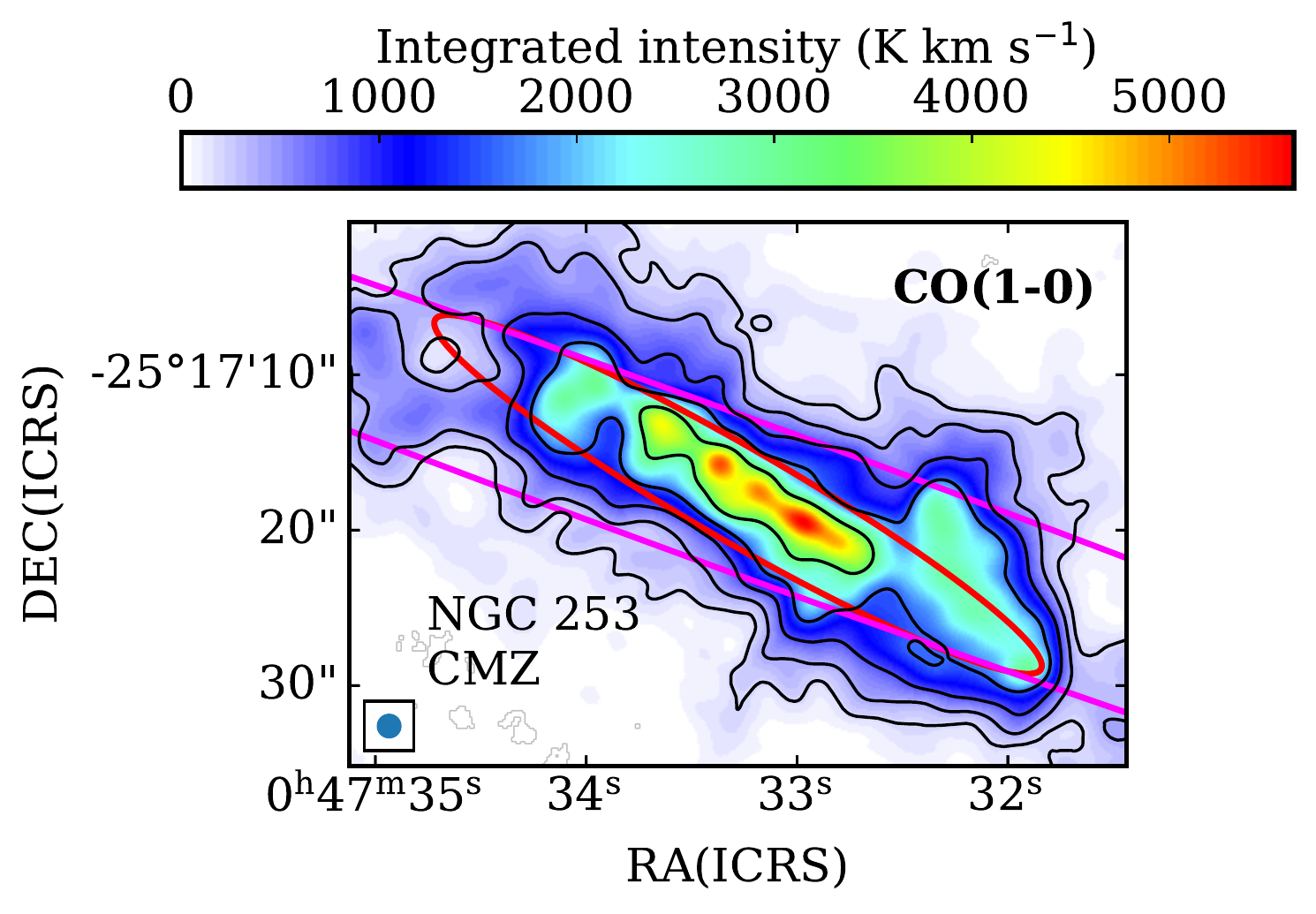}
}
\caption{(Left) The IRAC 8$\mu$m image is shown with contours, while the integrated-intensity image of CO(2-1) is shown with color in the logarithmic scale.
The contour levels are 25, 50, 100, 200, and 500 MJy/sr. Approximate positions of bar orbits ($x_1$ orbits) and inner orbits that form the CMZ ($x_2$ orbits) are shown with magenta and red ellipses, respectively. The beam size is shown as a blue ellipse on the left-bottom corner. The area shown in Figure \ref{fig:mom0} and the right panel are indicated with a white dashed rectangle in Figure \ref{fig:mom0} (left). For more detailed orbit models, we refer readers to other works \citep[e.g.,][]{2000PASJ...52..785S,2001ApJ...549..896D,2022arXiv220604700L}. (Right) The integrated intensity image of CO(1-0) overlaid with the same orbits as the left panel. \label{fig:orbit}}
\end{figure*}

In this paper, we report an enhancement of \hocop\ at the orbital intersections near the center of the starburst galaxy NGC~253. NGC~253 is one of the nearest \citep[$d=3.5$\,Mpc;][]{2005MNRAS.361..330R} and most studied starburst galaxies. Its nuclear ring forms a central molecular zone (CMZ) within a few hundred parsec scale, with a mass of $\sim 2\times 10^8\,M_{\odot}$ within a radius of $r = 150$\,pc \citep{leroy_alma_2015}. This large reservoir of molecular gas enables active star formation \citep{sakamoto_molecular_2006,sakamoto_star-forming_2011,bolatto_suppression_2013,krieger_molecular_2019,2020MNRAS.491.4573R}, which affects the properties of molecular gas \citep[e.g., through heating ][]{mangum_fire_2019}. The gas in the bar orbit ($x_1$ orbits) is being fed to the center of NGC~253 ($x_2$ orbits), and shocks are expected when this gas flow collides with the nuclear ring as discussed above for other galaxies. Figure \ref{fig:orbit}(left) shows IRAC 8$\mu$m \citep{2009ApJ...703..517D,slvl} and CO(2-1) integrated-intensity (ALMA data 2018.1.01321.S; PI: Faesi) images covering most of NGC~253 to indicate the locations of these orbits. The association between shocks at orbital intersections and the chemistry has been proposed by \citet{2000A&A...355..499G} and \citet{2015ApJ...801...63M}. 

At the center of NGC~253, a large number of molecular species are detectable \citep{2006ApJS..164..450M,2015ApJ...801...63M}. To fully explore the chemical complexity in this galaxy, we conducted the ALMA large program AL{\footnotesize MA} Comprehensive High-resolution Extragalactic Molecular Inventory \citep[ALCHEMI;][]{2021A&A...656A..46M}. ALCHEMI is a wide-frequency, unbiased spectral scan mosaic toward the CMZ of NGC~253 at a common 1.6$\arcsec$ resolution. This survey has discovered high cosmic-ray ionization rates \citep{2021A&A...654A..55H,2021ApJ...923...24H,2022arXiv220403668H}, detected a phosphorus-bearing species for the first time in an extragalactic source \citep{2022A&A...659A.158H}, and detected methanol masers \citep{2022arXiv220503281H} in the CMZ of NGC~253.  We utilize the ALCHEMI data to study multiple transitions of \hocop.

This paper is organized as follows. Section \ref{sec:obs} describes our observational parameters and data analysis, and images of integrated intensities are presented in Section \ref{sec:integint}. Derived column densities of \hocop\ are shown in Section \ref{sec:coldens}, while the \hocop/\cotw\ ratios are discussed using chemical models in Section \ref{sec:chem}. In Section \ref{sec:discussion}, we discuss our results including a comparison with ice abundances. Our results are summarized in Section \ref{sec:summary}.

\section{Observations and data analysis} \label{sec:obs}

The ALCHEMI spectral survey mosaic of the NGC~253 CMZ was performed between 2017 and 2019, with ALMA 12m-antenna and 7m-antenna arrays. It covered a broad frequency range between 84 and 373~GHz (Bands 3 to 7, avoiding deep atmospheric lines), down to sensitivities $\sim 10$~mK per 10\,km\,s$^{-1}$ channel.
The ALCHEMI data products have a uniform angular resolution of 1$\farcs$6 and cover a field of view of $50\arcsec\times20\arcsec$ centered on the CMZ of NGC~253 (phase center $\alpha=00^h47^m33.28^s$, $\delta=-25^\circ17'17.7''$ (ICRS)). The extent of the largest recoverable angular scale is greater than or equal to 15$\arcsec$. A detailed description of the ALCHEMI survey products can be found in \citet{2021A&A...656A..46M}.

We extracted the line cubes around the transitions of \hocop\ with a velocity resolution binned to 10\,km\,s$^{-1}$. Within the $\sim 290$~GHz coverage of the ALCHEMI survey, we find 14 detectable \hocop\ transitions, occurring every $\sim 21.4$~GHz. However, some transitions are severely blended with transitions from other species and are not used in this analysis. The spectroscopic parameters and spectral channel RMS values for the \hocop\ transitions used in this paper are listed in Table~\ref{tab:spec}.

\begin{deluxetable*}{ccccccc} 
\tablecolumns{7} 
\tablewidth{0pc} 
\tablecaption{\hocop\ Spectroscopic Properties and RMS Noise Values\label{tab:spec}}
\tablehead{\colhead{Transition} &\colhead{$\nu_{rest}^{(a)}$} &\colhead{$E_{\rm up}$$^{(b)}$} &\colhead{$log(A_{\rm ul})$$^{(c)}$} &\colhead{RMS$^{(d)}$} &\colhead{RMS$^{(e)}$} &\colhead{Blending$^{(f)}$}\\
\colhead{} & \colhead{(GHz)} &\colhead{(K)} &\colhead{(s$^{-1}$)} &\colhead{(mJy beam$^{-1}$)} &\colhead{(mK)} &\colhead{}}
\startdata 
 4$_{0,4}$-3$_{0,3}$ &85.531&10.3&-4.63&0.19&12.&Potential blending with CH$_3$CCH. Minor blending with U-lines\\
 5$_{0,5}$-4$_{0,4}$&106.914&15.4&-4.33&0.20&8.4 & N/A\\
 6$_{0,6}$-5$_{0,5}$&128.295&21.6&-4.08&0.33&9.5 &Minor blending with U-lines\\
 7$_{0,7}$-6$_{0,6}$&149.676&28.7&-3.88&0.42&9.0 &Potential blending with U-lines\\
 8$_{0,8}$-7$_{0,7}$&171.056&36.9&-3.70&0.75&12. &Potential blending with CH$_3$CCH.\\
 12$_{0,12}$-11$_{0,11}$&256.566&80.0&-3.16&0.99&7.2 &Potential blending with CH$_3$CCH, HC$_3$N $v7=2$\\
 \hline
 \multicolumn{7}{c}{Transitions below were not used for analysis}\\
 \hline
 9$_{0,9}$-8$_{0,8}$ &192.435&46.2&-3.54& &&Blended with U-line\\
 10$_{0,10}$-9$_{0,9}$ &213.813&56.4&-3.40& &&Blended with C$_2$H$_5$OH\\
 11$_{0,11}$-10$_{0,10}$ &235.190&67.7&-3.28&&&Blended with SO$_2$\\
 13$_{0,13}$-12$_{0,12}$ &277.941&93.4&-3.06&&&Blended with U-line\\
 14$_{0,14}$-13$_{0,13}$ &299.314&107.7&-2.96&&&Non detection\\
15$_{0,15}$-14$_{0,14}$ &320.686&123.1&-2.87&&&Non detection\\
16$_{0,16}$-15$_{0,15}$ &342.056&139.6&-2.78&&&Non detection\\
17$_{0,17}$-16$_{0,16}$ &363.424&157.0&-2.70&&&Non detection\\
\enddata 
\tablecomments{$(a)$ Rest frequency; $(b)$ Upper level energy of the transition; $(c)$ $A_{ul}$: Einstein coefficient of spontaneous emission. All values were taken from the Cologne Database for Molecular Spectroscopy 
\citep[CDMS; https://cdms.astro.uni-koeln.de;][]{2001AA370L49M,2005JMoSt.742..215M,2017AandA...602A..34B}; (d) and (e) RMS values of a single channel with $\Delta v=10$\,km\,\ps\ in mJy beam$^{-1}$ and mK units; (f) presence of blending; Transitions are shown with quantum numbers $J_{K_a, K_c}$. Only $K_a = 0$ transitions are shown because transitions with $K_a \neq 0$ are not detected due to their higher energy state and lower Einstein coefficients. The upper part of this table shows transitions used for the analyses of this paper, whose line shapes are separable from neighboring lines. The lower part lists transitions with severe blending or without reliable detection. ``Potential blending" means the case where the line centers are separated by more than 200\,km\,s$^{-1}$, but the line wing can contaminate the moment maps of \hocop\ transitions.} 
\end{deluxetable*} 

\section{Integrated intensities}\label{sec:integint}
\begin{figure*}[h]
\centering{
\includegraphics[width=0.99\textwidth]{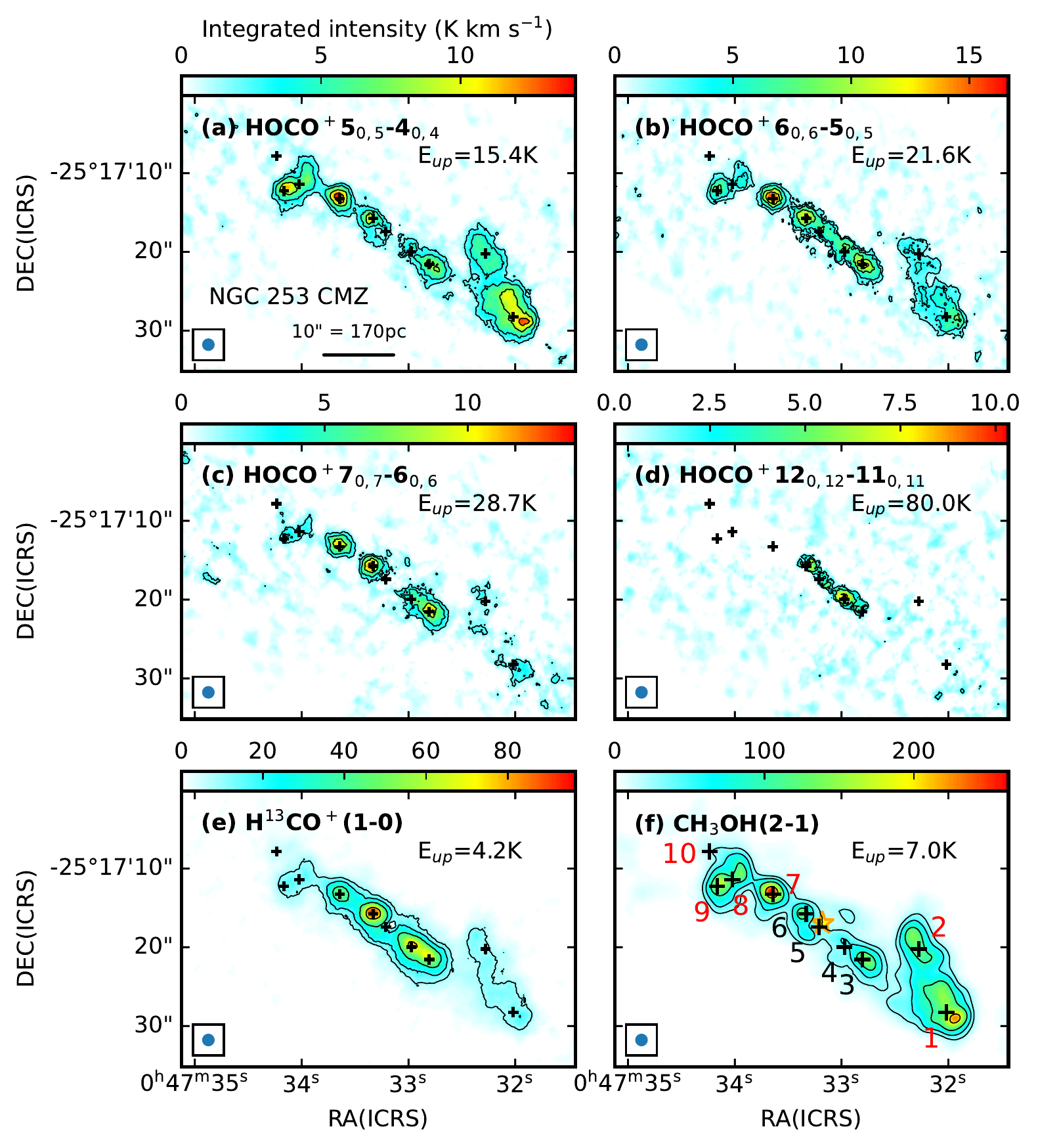}
}
\caption{Velocity-integrated intensity maps of (a-d) four \hocop\ transitions, (e) \httcop(1-0), and (f) \methanol($2_k-1_k$). Contour levels are (a-d) [2, 4, 8, 12], (e) [9.0, 18.0, 36.0, 72.0], and (f) [25.0, 50.0, 100.0, 200.0] K km s$^{-1}$. The beam size of $1.6\arcsec \times 1.6\arcsec$ ($27\,{\rm pc}\times 27\,$pc) is shown at the bottom left corner in each panel as a blue circle. Locations of GMCs identified by \citet{leroy_alma_2015} are shown as black crosses and labeled in panel(f). Coordinates of these GMCs are given in Table \ref{tab:gmc}. The kinematic center known as ``TH2" \citep{turner_1_1985} with the revised coordinate by \citet{2020MNRAS.493..627C} is shown as a yellow star with orange edges in panel (f). GMCs with Class I methanol maser detection by \citet{2022arXiv220503281H} are labeled with red numbers in panel (f).\label{fig:mom0}}
\end{figure*}

Figure \ref{fig:mom0} (a-d) shows the velocity-integrated intensity (moment 0) images of \hocop\ in multiple transitions (see Table \ref{tab:spec} for their properties). Many transitions used in this work have neighboring lines, and it is not possible to make the moment 0 images simply by collapsing neighboring channels. To exclude this contamination, we applied the 3-D mask made from the position-position-velocity space only including pixels with CO $J=1-0$ detections above the $20\sigma$ level. The 20$\sigma$ cutoff may sound unnecessarily high, but the signal-to-noise ratios of \hocop\ transitions are more than 100 times lower than that of CO(1-0). Therefore, this mask does not exclude any notable \hocop\ emission but helps to exclude the contamination from nitrogen sulfide transitions neighboring with CO(1-0) to be included in the mask. Despite the elimination of contamination with this mask, the only transition that is free from contamination is $5_{0,5}-4_{0,4}$ (panel a) of Figure \ref{fig:mom0}. Other images that are relatively less affected from blending are also shown in Figure \ref{fig:mom0} (panels b-d). The level of contamination is usually very low ($<10\%$) except for GMC 5, where there is little \hocop\ emission and stronger emission from neighboring lines. Transitions $4_{0,4}-3_{0,3}$ and $8_{0,8}-7_{0,7}$ can still be used to obtain column densities as we use spectral fitting, but their images are not shown. A low-excitation line of \hocop\ ($5_{0,5}-4_{0,4}$; $E_{\rm up}=15.4\,$K) shows peaks near the outer CMZ, in giant molecular clouds (GMCs) 1, 7, and 9 \citep[GMC numbering is shown in panel f;][see also Appendix \ref{sec:app_gmc} for coordinates]{leroy_alma_2015}.
On the other hand, the higher-excitation transitions ($E_{\rm up}\gtrsim 30\,$K) peak closer to the center (GMCs 3 and 6). 

Panels (e) and (f) of Figure \ref{fig:mom0} show integrated-intensity images of \methanol($2_k-1_k$) (group of transitions at $\nu_{rest} \sim 96.74$ GHz with the strongest transition $2_0^+ - 1_0^+$)\footnote{These transitions may not be in local thermodynamic equilibrium (LTE), but are ``quasi thermal" and they are not identified as masing \citep{2022arXiv220503281H}.} and H$^{13}$CO$^+$(1-0) ($\nu_{rest} = 86.75$ GHz) for comparison. The emission distribution of \methanol($2_k - 1_k$) is similar to that of the low-$J$ transitions of \hocop, as well as low-$J$ transitions of HNCO, and SiO \citep[][Huang et al., in preparation]{2015ApJ...801...63M}. On the other hand, the distribution of H$^{13}$CO$^+$(1-0), which is rather similar to that of molecules with strongest emission \citep[e.g., CO, HCO$^+$, HCN, CS, etc.][]{2015ApJ...801...63M, 2021A&A...656A..46M}, is clearly different from that of \hocop. The H$^{13}$CO$^+$(1-0) emission is concentrated near the center of NGC 253 (GMCs 3-7) instead of the outer CMZ.

\begin{figure*}[h]
\centering{
\includegraphics[width=0.99\textwidth]{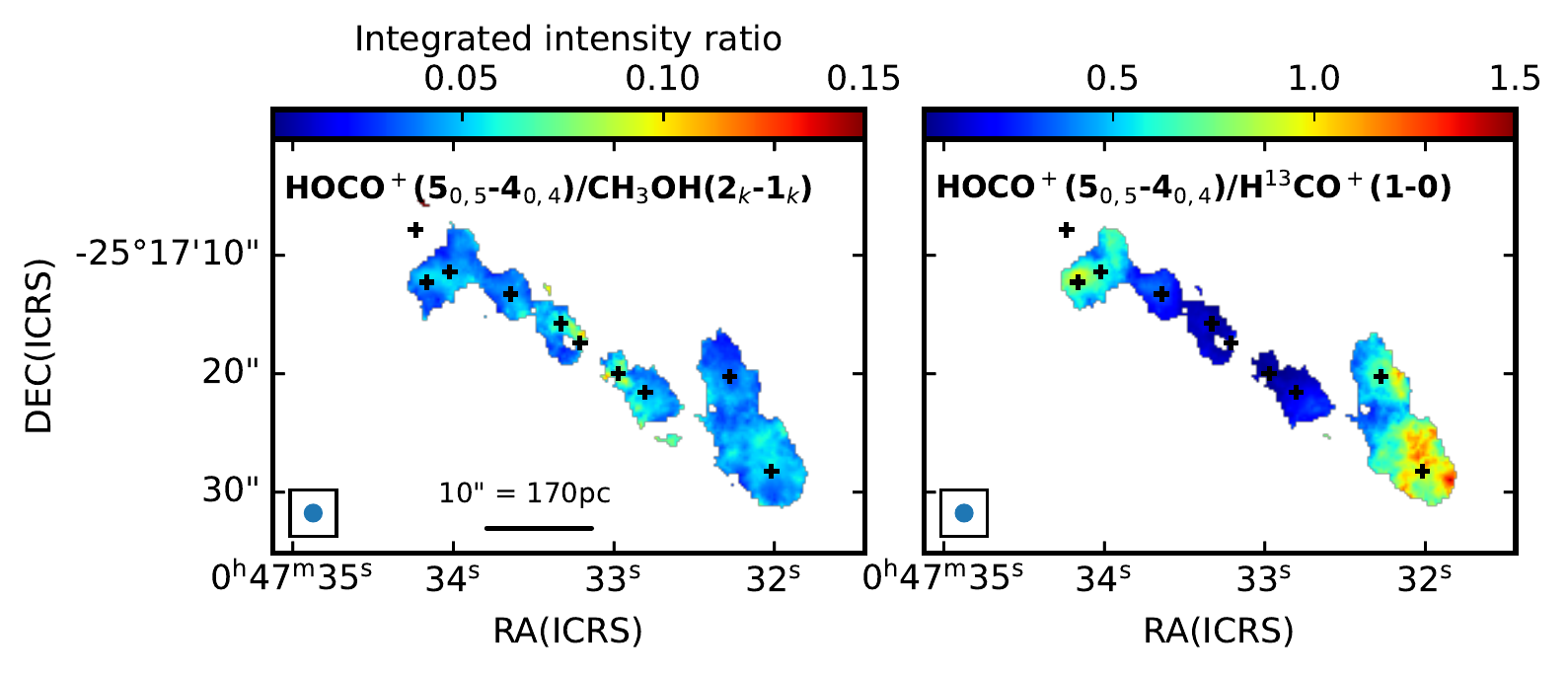}
}
\caption{Integrated intensity ratios of (left) \hocop($5_{0,5}-4_{0,4}$)/CH$_3$OH ($2_k-1_k$) and (right) \hocop($5_{0,5}-4_{0,4}$)/H$^{13}$CO$^+$(1-0). Only pixels with integrated intensity of $>3\sigma$ for each transition and ratio are shown. Ranges of color scales are set to be 0.003-0.15 (left) and 0.03-1.5 (right), respectively. Because the maximum/minimum ratios of color scales are 50 for both images, variations in color scales between the left and right panels represent variations in the \hocop($5_{0,5}-4_{0,4}$)/CH$_3$OH ($2_k-1_k$) and \hocop($5_{0,5}-4_{0,4}$)/H$^{13}$CO$^+$(1-0) ratios. \label{fig:ratio}}
\end{figure*}

The similarity between integrated intensities of \hocop\ ($5_{0,5}-4_{0,4}$) and CH$_3$OH ($2_k-1_k$) and the difference between those of \hocop\ ($5_{0,5}-4_{0,4}$) and H$^{13}$CO$^+$(1-0) are highlighted in Figure \ref{fig:ratio}. While the \hocop($5_{0,5}-4_{0,4}$)/CH$_3$OH ($2_k-1_k$) ratios are relatively constant, the \hocop($5_{0,5}-4_{0,4}$)/H$^{13}$CO$^+$(1-0) ratio varies significantly. All these transitions have relatively low upper state energies (\hocop\ ($5_{0,5}-4_{0,4}$): $E_u=15.4\,$K, CH$_3$OH ($2_0^+-1_0^+$):$E_u=7.0\,$K, H$^{13}$CO$^+$(1-0): $E_u=4.2\,$K), and these ratio maps should be good proxies for variations in column densities of these molecules.

\section{Column densities and fractional abundances}\label{sec:coldens}

\begin{figure*}[h]
\centering{
\includegraphics[width=0.49\textwidth]{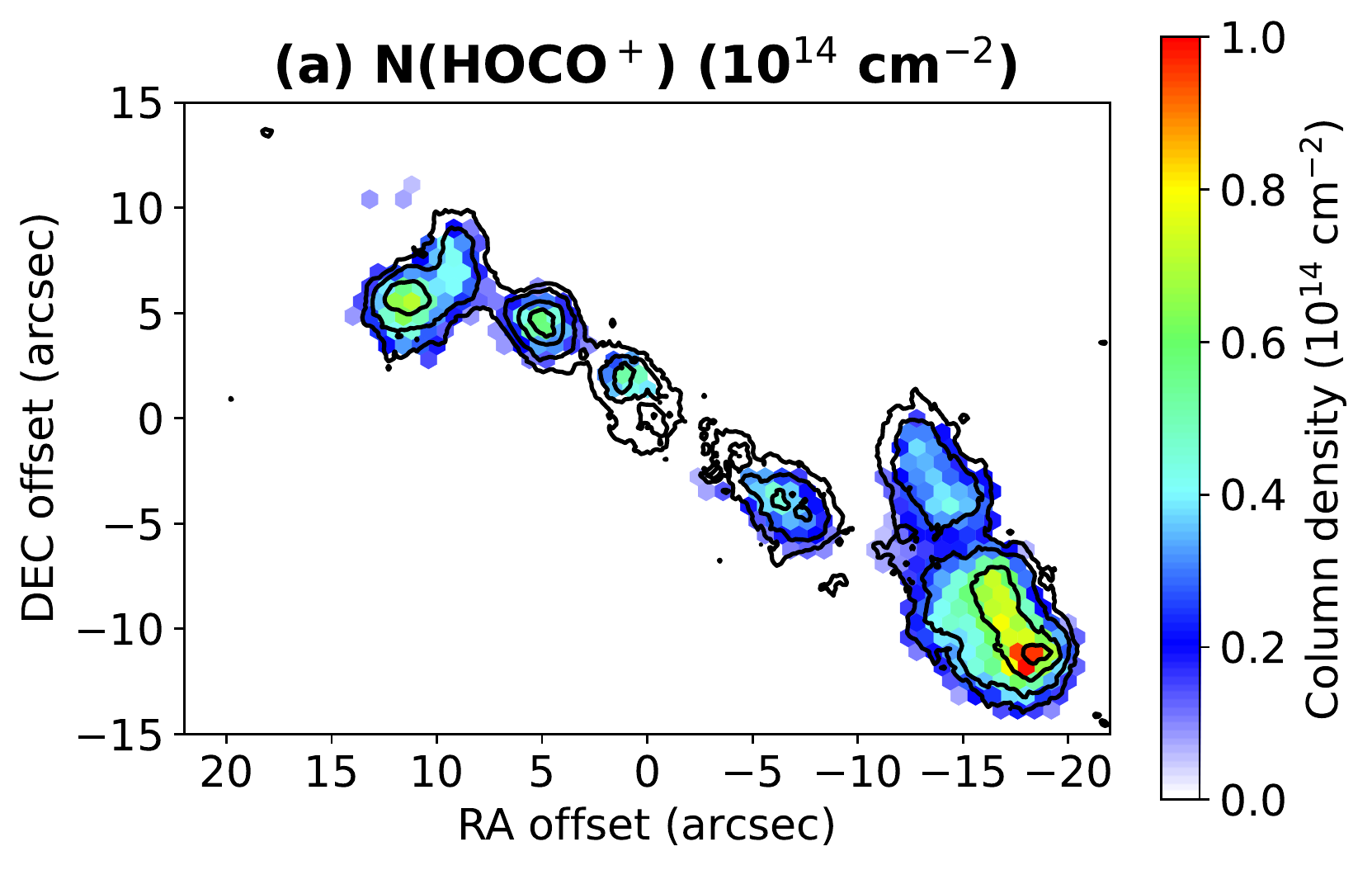}
\includegraphics[width=0.49\textwidth]{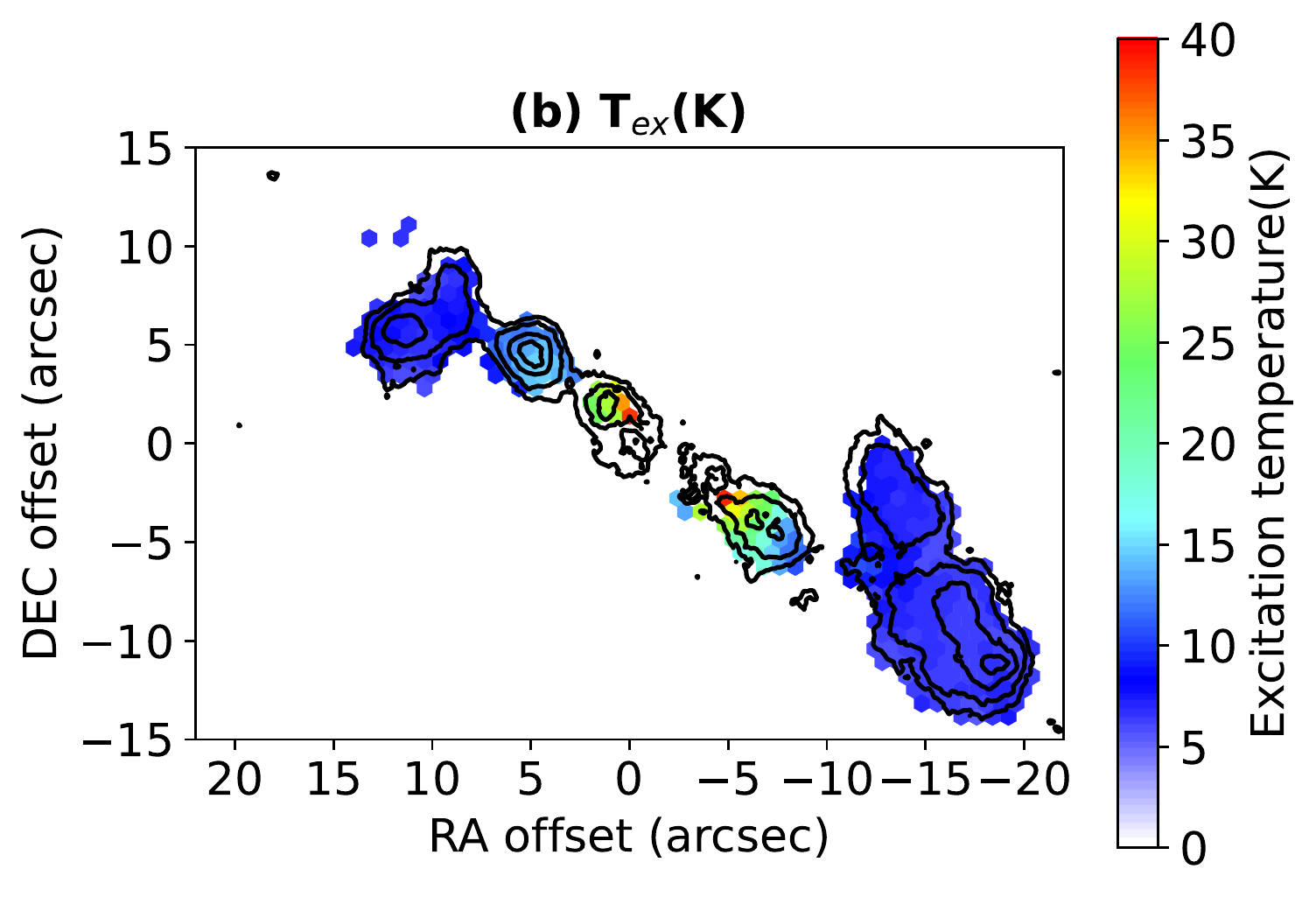}}
\centering{
\includegraphics[width=0.49\textwidth]{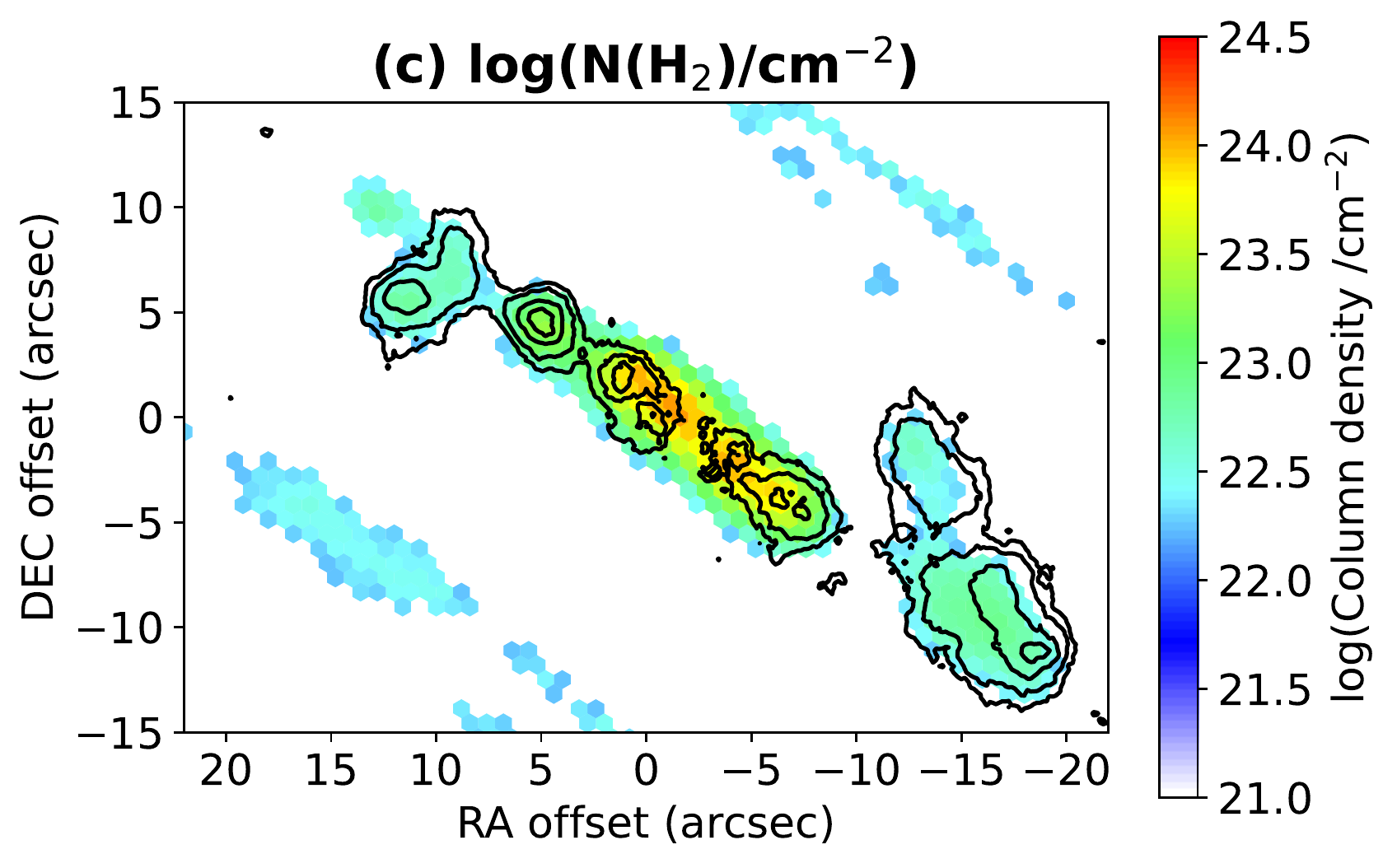}
\includegraphics[width=0.49\textwidth]{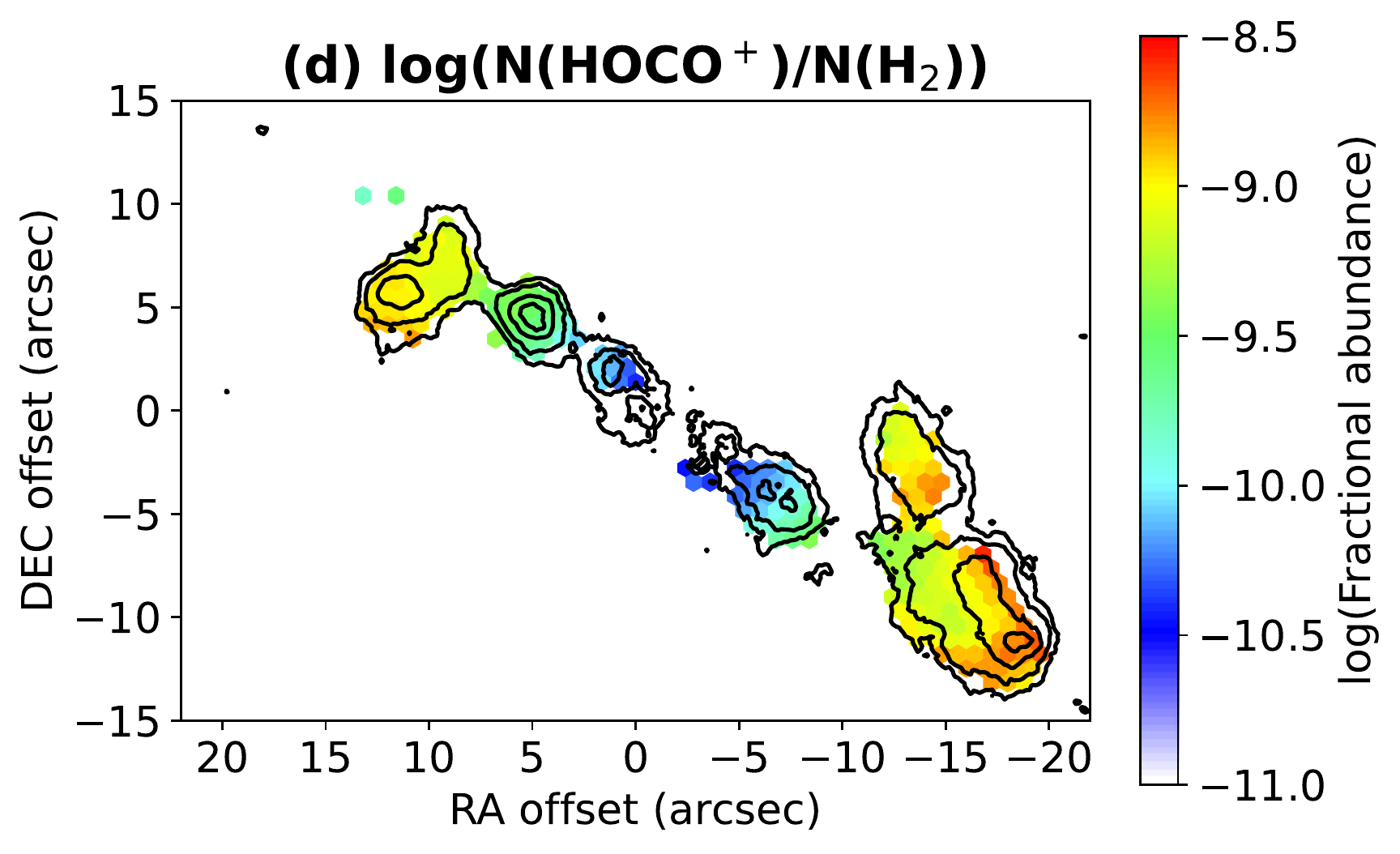}
}
\caption{Maps of (a) the \hocop\ column density, (b) the excitation temperature of \hocop derived from CASSIS, (c) the total  H$_2$ column density derived from the dust continuum shown in a logarithmic scale, and (d) the fractional abundance of \hocop\ shown in the logarithmic scale. The black contours show \hocop\ ($5_{0,5}-4_{0,4}$) integrated intensities as in Figure \ref{fig:mom0}. \label{fig:hocop_col}}
\end{figure*}

\begin{figure*}[h]
\centering{
\includegraphics[width=0.49\textwidth]{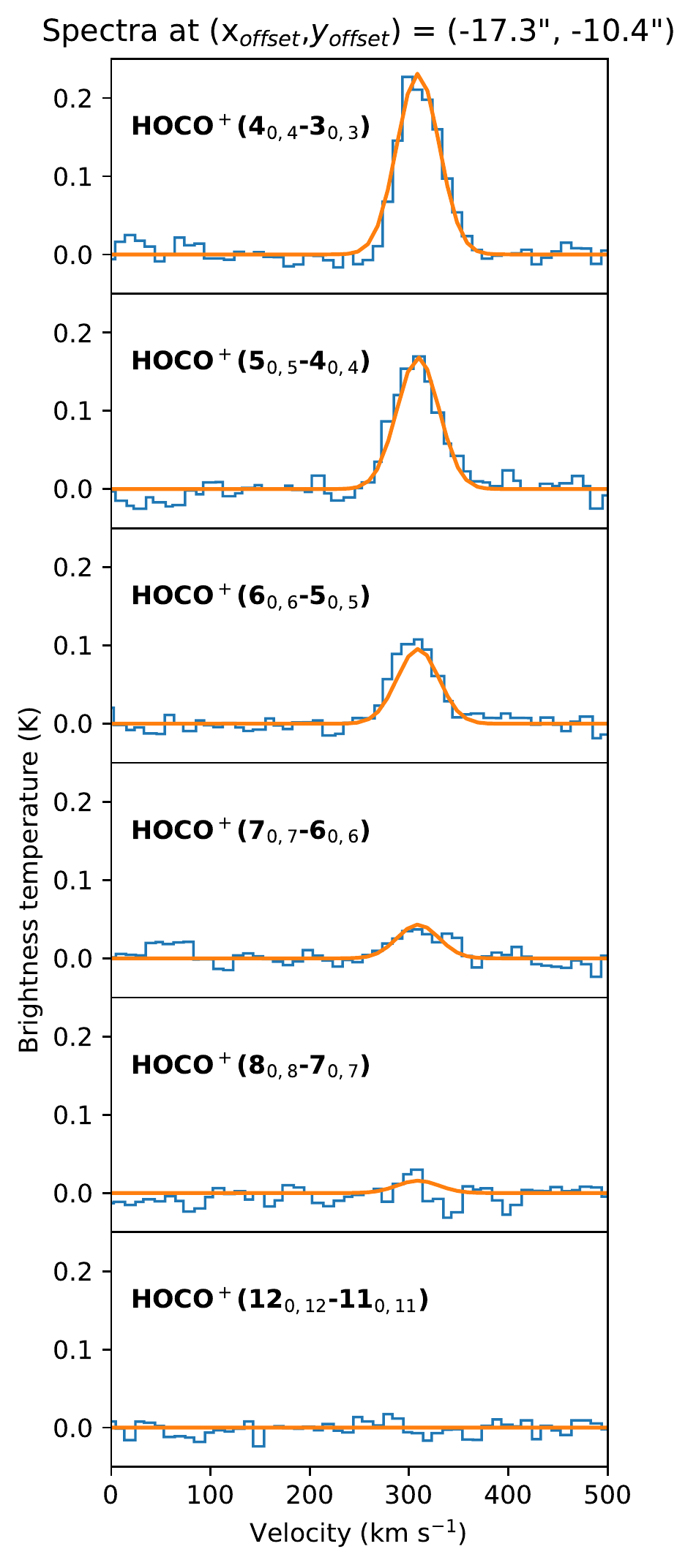}
\includegraphics[width=0.468\textwidth]{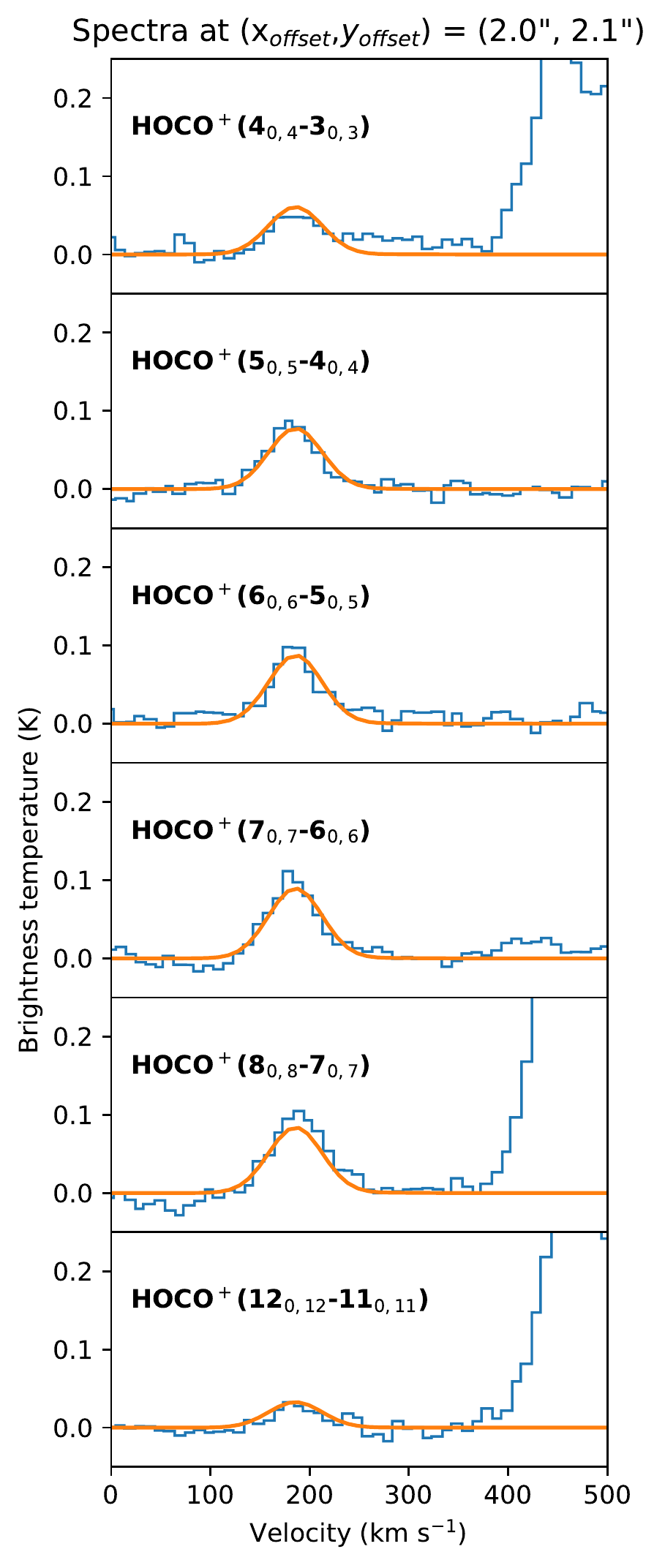}}
\caption{Examples of observed spectra (blue) versus modeled spectra by CASSIS (orange) at ($x_{\rm offset}$,$y_{\rm offset}$) of (Left Panel) (-17.3$\arcsec$,-10.4$\arcsec$) and (Right Panel) (2.0$\arcsec$,2.1$\arcsec$) relative to the phase center of our observations. \label{fig:spectra}}
\end{figure*}

Figure \ref{fig:hocop_col}(a) shows the column density map of \hocop. These column densities were derived using the public software CASSIS \citep[http://cassis.irap.omp.eu/][]{2015sf2a.conf..313V} supplied with spectroscopic constants from the spectroscopic database from the CDMS \citep{2001AA370L49M,2005JMoSt.742..215M}.
CASSIS calculates molecular column densities based on input spectral line brightness temperatures with consideration of optical depths, either with an LTE assumption or, if collisional rates are available, non-LTE assumption. We used a Markov Chain Monte Carlo (MCMC) algorithm assuming LTE to fit column densities, excitation temperatures, line velocities, and line widths (see Appendix \ref{sec:app_cassis}). Column densities were calculated only for the pixels with $>3\sigma$ detection of both \hocop($4_{0,4}-3_{0,3}$) and \hocop($5_{0,5}-4_{0,4}$) at velocities within 30\,\kms\ from the line center. Line-center velocities used for this 3-$\sigma$ detection criterion are determined from the image cube of CO(1-0) from ALCHEMI data. These line-center velocities from CO may be different from \hocop\ velocities fitted from CASSIS. Instead of deriving the column densities and excitation temperatures on a pixel-by-pixel basis, we bin the intensities within hexagonal pixels with a horizontal length of 0\farcs8, half of the image spatial resolution of 1.6\arcsec, to reduce the computational time running CASSIS.  Examples of spectral fitting are shown for hexagonal pixels located at ($x_{\rm offset}$,$y_{\rm offset}$)=($-17.3\arcsec$,$-10.4\arcsec$) and ($2.0\arcsec$,$2.1\arcsec$) in Figure \ref{fig:spectra} where the offset is taken from the phase center. In general, the observed spectra fit well with the LTE spectra. The resulting column-density distribution appears similar to that of the moment~0 images of low-excitation transitions (e.g., Figure~\ref{fig:mom0}a). 

Excitation temperatures derived from the above spectral fitting are shown in Figure~\ref{fig:hocop_col}(b). While regions far from the kinematic center\footnote{The kinematic center of NGC~253 is located near GMC~5 \citep{turner_1_1985,2010ApJ...716.1166M}. Although there is a debate on the exact location of the kinematic center, the difference of 0\farcs7 appearing in the literature does not affect our discussion.} of NGC~253 show low excitation temperatures of $\lesssim 10\,$K (GMCs 1, 2, 8, and 9), the excitation temperatures become higher, up to 40 K, near the center of NGC~253 (GMCs 3 and 6). 

We also obtain the total hydrogen column densities (Figure \ref{fig:hocop_col}c) to calculate fractional abundances of \hocop\ (the \hocop\ column densities divided by total hydrogen column densities N(H$_2$)). The total H$_2$ column densities were derived from the dust continuum image at 361.5 GHz shown by \citet{2021ApJ...923...24H} with the derivation method based on \citet{1983QJRAS..24..267H} for pixels above $3\sigma$ detection. A simplified formula is given as Equation (3) in \citet{mangum_fire_2019}:
\begin{equation}
    N(H_2) ({\rm cm^{-2}}) \sim 7.0\times 10^{22} R_{dg} \left (\frac{\lambda ({\rm mm})}{0.4}\right)^\beta \frac{T_R({\rm K})}{T_d ({\rm K})},
\end{equation}
where $N(H_2)$ is the molecular hydrogen column density, $R_{dg}$ is the dust-to-gas mass ratio, $\lambda$ is the wavelength, $T_R$ is the radiation temperature, and $T_d$ is the dust temperature. This formula is valid for $h\nu\ll kT_d$ (the Rayleigh-Jeans approximation). We use the emissivity $\beta =1.5$, a dust temperature $T_{\rm dust}=30\,$K, and a dust-to-gas mass ratio of 150 following \citet{mangum_fire_2019}. This estimate of the dust temperature is close to the observed value, but some dust components may be warmer. \citet{2018ApJ...860...23P} derived dust temperature components of 37, 70, and 188\,K in the central region of NGC 253 using their assumed source size of $17.3\arcsec \times 9.2\arcsec$ from their Herschel and SOFIA observations. These components contain mass fractions of 65, 26, and 9\,\%, respectively. If the dust is warmer, the actual column density should be smaller by a similar factor; for instance, a factor of 5 smaller if $T_{\rm dust} = 150\,$K. The column-density dependence on the dust temperature becomes larger than $\propto \frac{1}{T_d}$ when the dust is cold ($T_d \lesssim 20\,$K) and one cannot use the Rayleigh-Jeans approximation, but we expect that there is a very small amount of cold dust in the center of NGC 253. 

The fractional abundance of HOCO$^+$ is higher at larger distances from the center of NGC~253, and it decreases by more than an order of magnitude at the center (Figure~\ref{fig:hocop_col}d). At peaks of \hocop (GMCs 1, 8, and 9), the fractional abundance is $\sim$\fnum{(1-2)}{-9}, similar to those observed in Galactic center clouds: $(2-8)\times10^{-9}$ \citep{minh_abundance_1991,2015MNRAS.446.3842A}. On the other hand, it is  orders of magnitude higher than those observed in Galactic disk clouds, which range from \enum{-13} to \fnum{5}{-11} \citep{vastel_abundance_2016,fontani_protonated_2018,majumdar_detection_2018}.

\section{\hocop/\cotw\ ratios}\label{sec:chem}
\begin{figure*}[h]
\centering{
\includegraphics[width=0.49\textwidth]{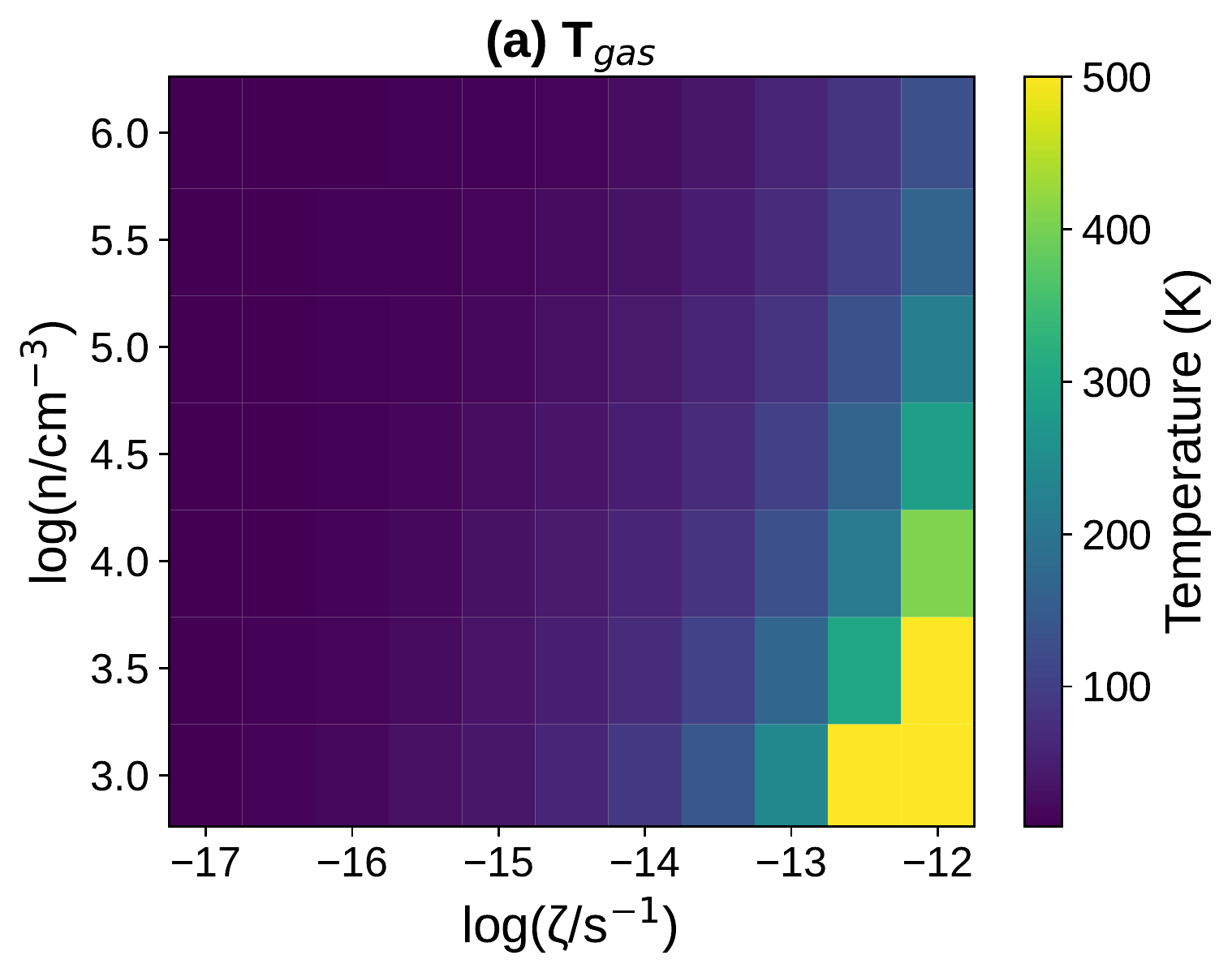}
\includegraphics[width=0.49\textwidth]{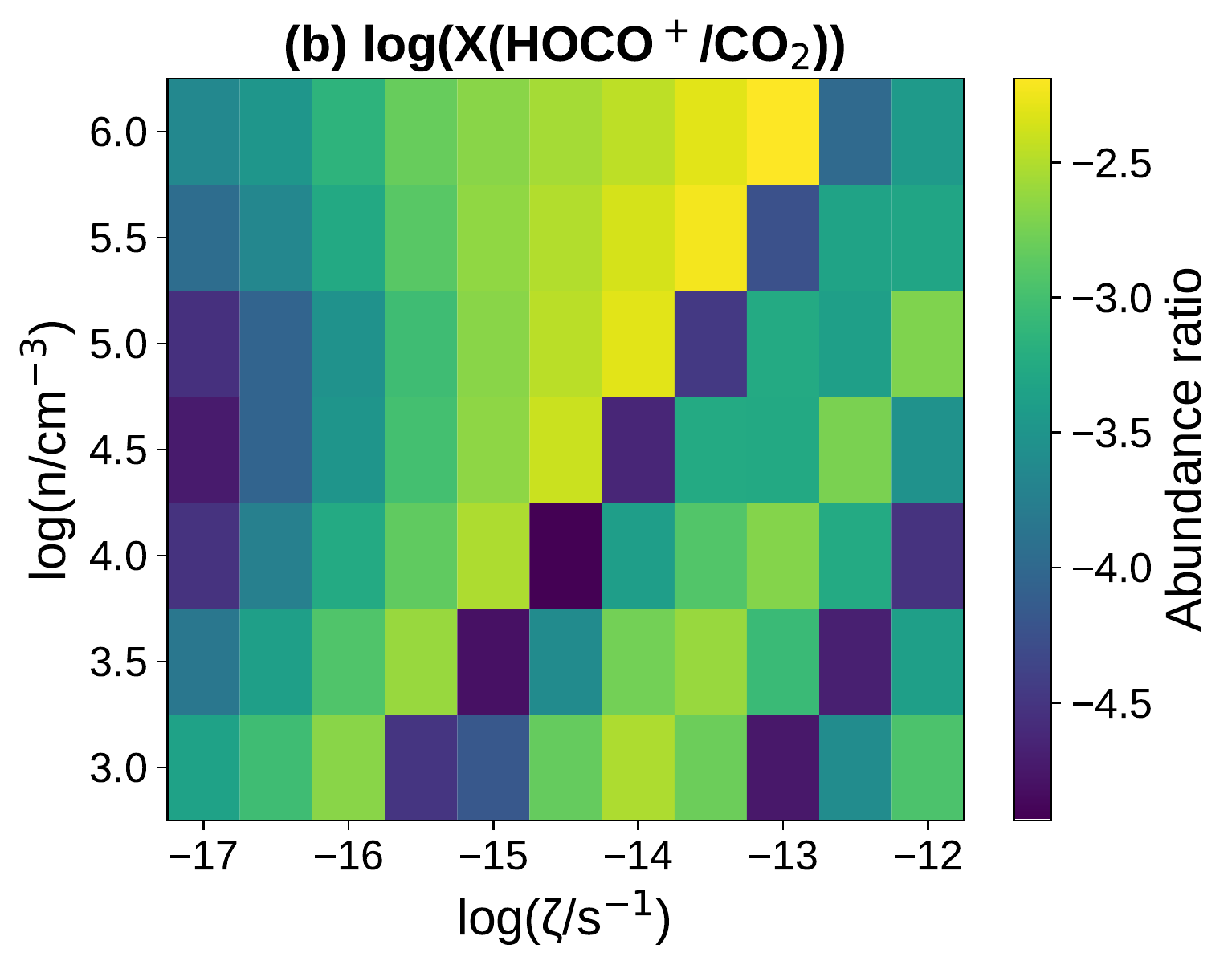}
}
\centering{
\includegraphics[width=0.49\textwidth]{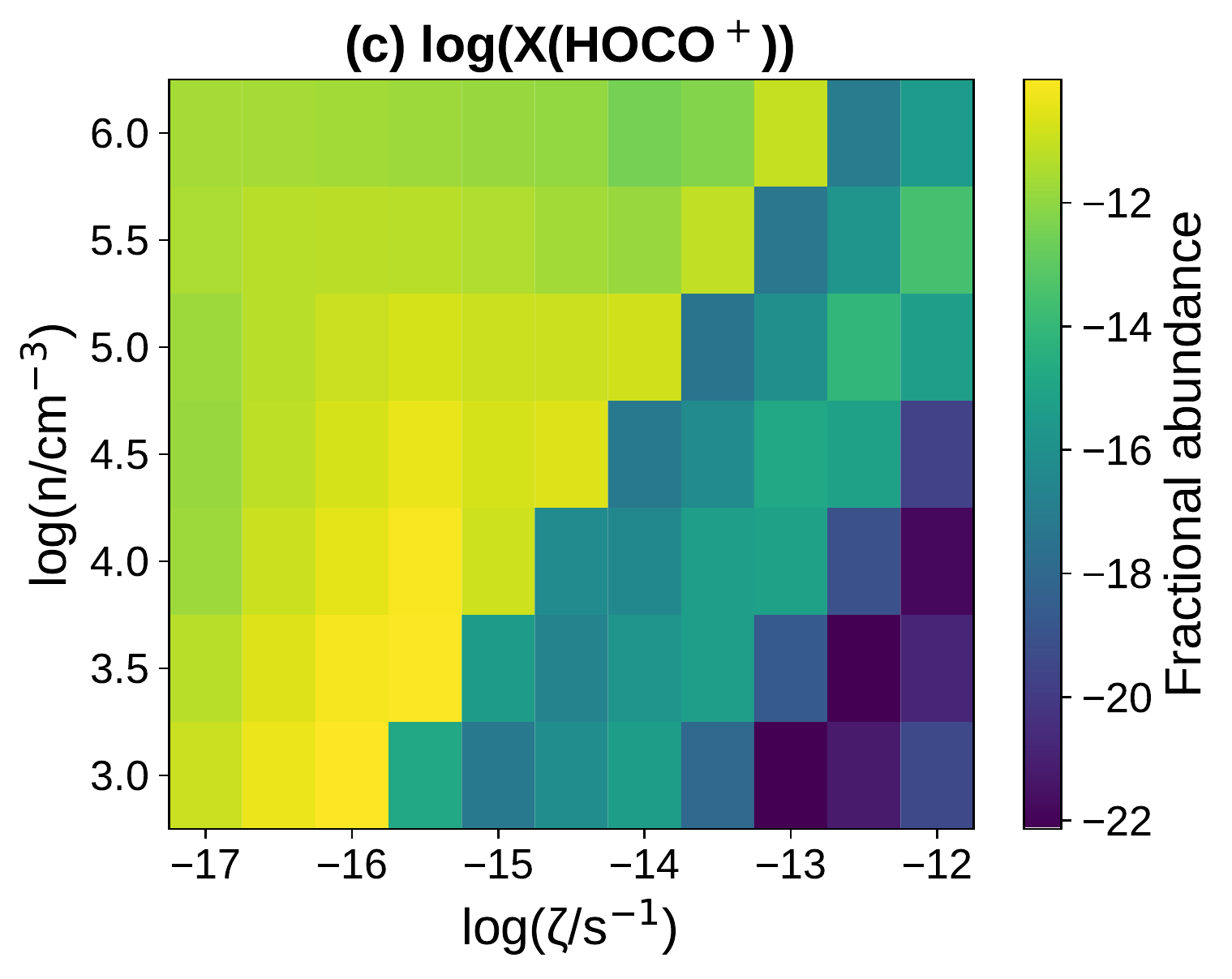}
\includegraphics[width=0.49\textwidth]{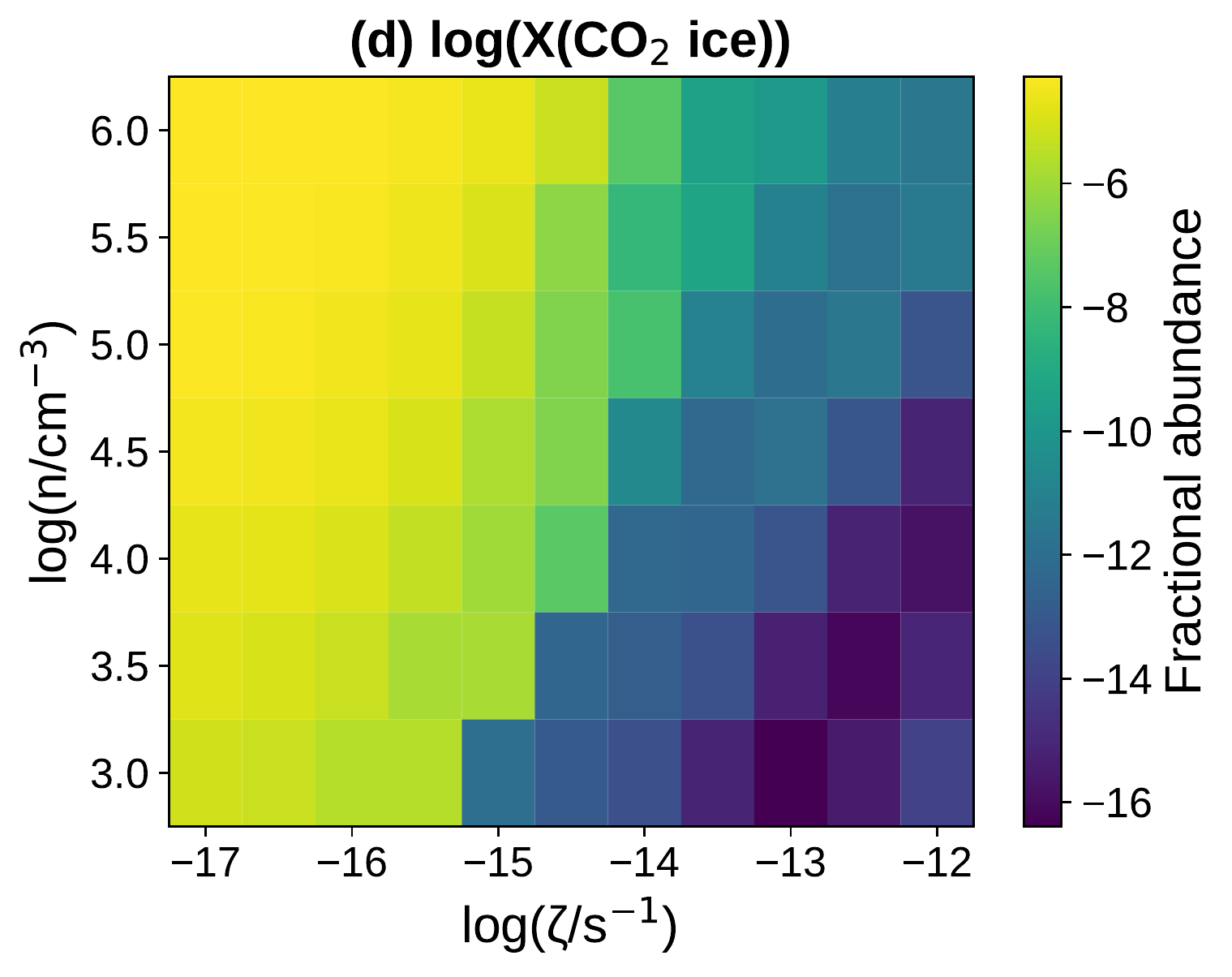}
}
\centering{
\includegraphics[width=0.49\textwidth]{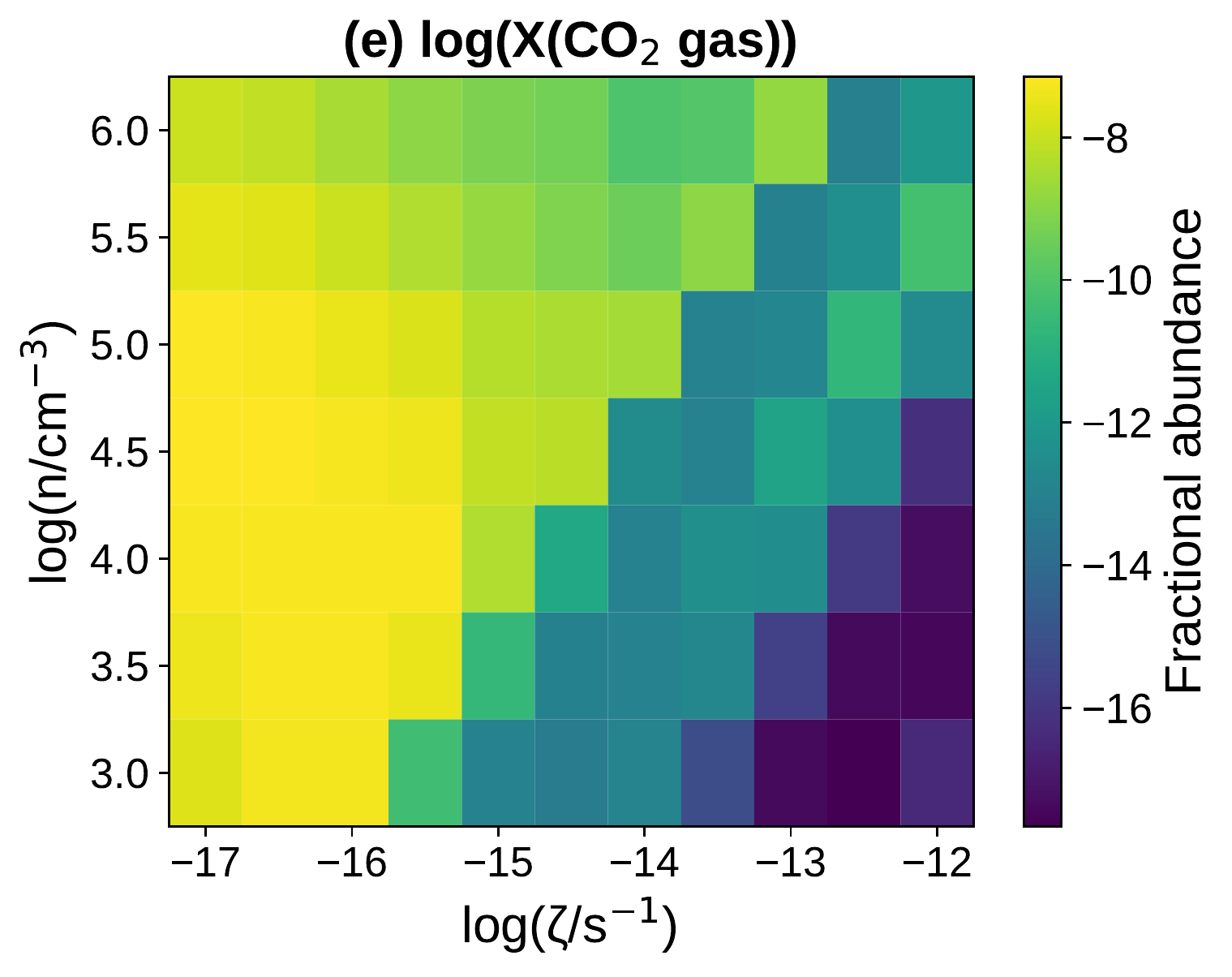}
\includegraphics[width=0.49\textwidth]{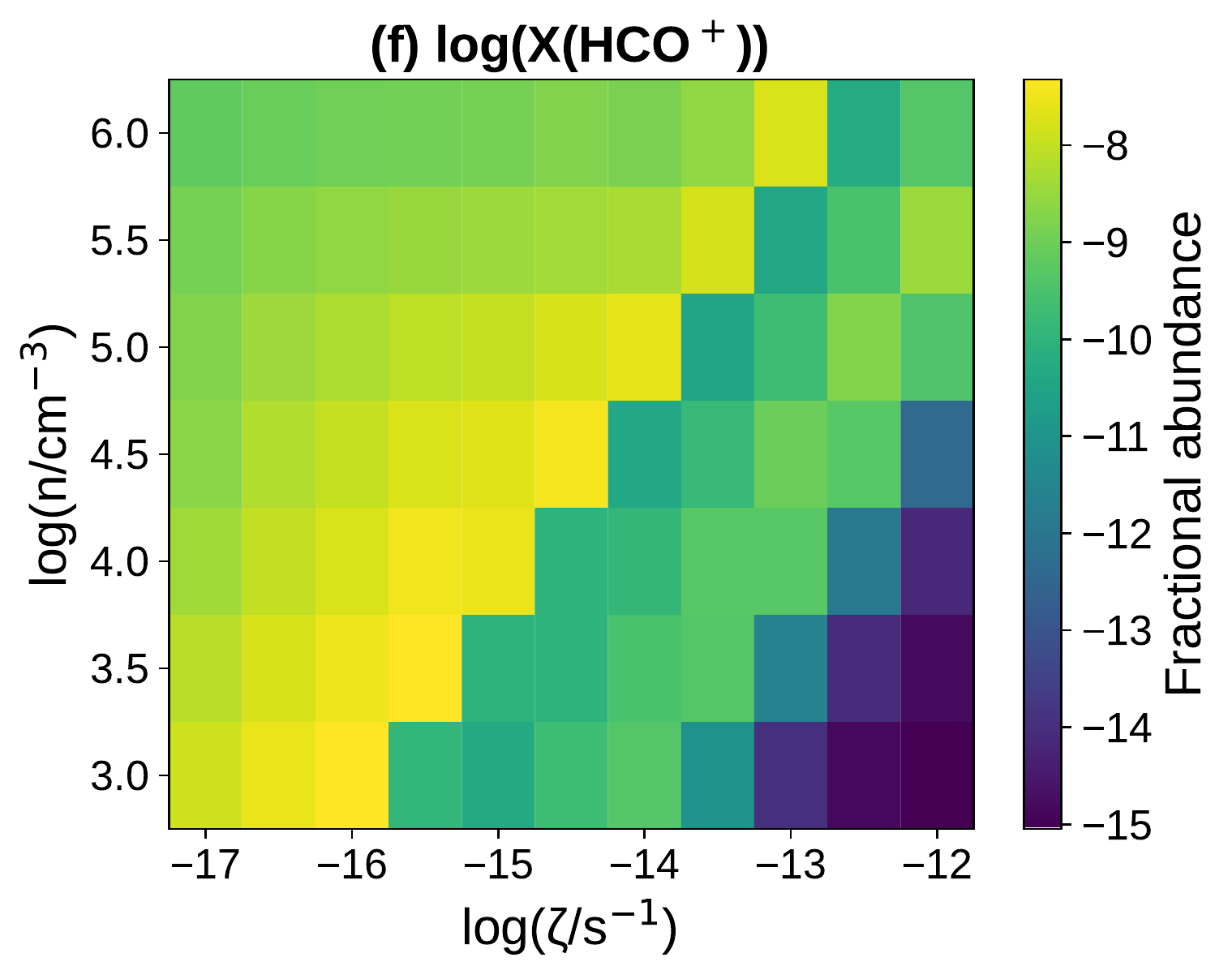}
}
\caption{Following modeled quantities are shown as functions of the density and cosmic-ray ionization rate: (a) gas temperatures from the Meudon code used for Nautilus chemical modeling (b) abundance ratios \hocop/\cotw\ (c) fractional abundances of \hocop, (d) \cotw\ ice (surface + mantle), e) \cotw\ gas, and (f) HCO$+$}. \label{fig:chem_model}
\end{figure*}

To estimate the gas-phase abundances of \cotw\ from \hocop, we ran chemical abundance models based on Nautilus \citep{2016MNRAS.459.3756R}, accounting for gas, ice surface, and ice mantle phases. In addition to the thermal evaporation, desorption from dust heating due to cosmic rays \citep{1993MNRAS.261...83H} is included in the model. Desorption through cosmic-ray heating of dust is where the dust grain is temporarily heated to a certain maximum temperature for a very short time scale ($\sim 10^{-5}\,$s), then cools down.
The model also includes photodesorption \citep{2009A&A...496..281O,2009ApJ...693.1209O} both from direct UV photons and cosmic-ray-induced UV photons with a default yield of \enum{-4} for all the grain species. We also ran a model with a desorption yield of \enum{-3} for \cotw. Our models do not include shocks.  We calculated grid models with varying densities ($n=10^3 - 10^6$\,cm$^{-3}$) and cosmic-ray ionization rates ($\zeta = 10^{-17} - 10^{-12}$\,\ps) following a similar approach as \citet{2021ApJ...923...24H}. Temperatures were calculated in the Meudon Photodissociation Region (PDR) code (ver. 1.5.4) \citep{2006ApJS..164..506L} (Figure \ref{fig:chem_model}a), and were fed to Nautilus to run chemical abundance models with a larger chemical network. Despite the high gas temperature with the high cosmic-ray ionization rate ($T>1000\,$K when $\zeta=10^{-12}$\,\ps\ and $n=10^3$\,cm$^{-1}$), the dust temperature calculated from the Meudon code remains cold, around 11\,K. We adopted a maximum visual extinction $A_{\rm V}=20\,$mag with a turbulent velocity of $1\,$km~s$^{-1}$, and used the temperature in the model at $A_{\rm V}=10\,$mag (Figure \ref{fig:chem_model}d), where the effects of the PDRs are negligible. We note that, unlike in the description of observational results, fractional abundances are expressed as abundances of certain species over total hydrogen abundances ($N_{\rm Htotal} = N_{\rm Hatom} + 2N_{H2}$), instead of molecular hydrogen abundances.

Figure~\ref{fig:chem_model} (b) shows \hocop/\cotw\ abundance ratios in the gas phase, that vary between $10^{-5} - 10^{-2}$ for the most part. Previous ALCHEMI studies have suggested that the cosmic-ray ionization rates in NGC~253 are at least a few orders of magnitude higher than that in the Galactic spiral arm clouds \citep{2021A&A...654A..55H,2021ApJ...923...24H,2022arXiv220403668H}, which increases the \hocop/\cotw\ abundance ratios due to an increased H$_3^+$ abundance. Although the cosmic-ray ionization rates are expected to be high, extremely high rates ($\zeta \gtrsim 10^{-13}$\,\ps\ for $n=10^5\,$cm$^{-3}$) would destroy \hocop\ (Figure \ref{fig:chem_model}c), especially in lower density regions. Therefore, we consider \hocop/\cotw\ ratios of $\sim 10^{-3} - 10^{-2}$ in later sections, which are taken from the parameter space where the \hocop\ abundance is moderately high\footnote{We note that the presence of \hocop\ is still consistent with the cosmic-ray ionization rates derived by \citet{2021A&A...654A..55H} and \citet{2021ApJ...923...24H} ($\zeta \gtrsim 10^{-14}$\,\ps\ for $n=10^5\,$cm$^{-3}$), but the value obtained by \citet{2022arXiv220403668H} ($\zeta \sim 10^{-13}$\,\ps\ for $n=10^5\,$cm$^{-3}$) would not allow high fractional abundances of \hocop.}.

It is worth noting that cosmic-ray-induced desorption mechanisms do not change the abundances significantly. This is because cosmic-ray-induced photodissociation is a more efficient form of destruction than desorption if the desorption yield is on the order of \enum{-3} or lower. According to \citet{2009A&A...496..281O}, the photodesorption yield of \cotw\ cannot go higher than a few times \enum{-3}, which means photodesorption is not significant. Desorption of \cotw\ due to cosmic-ray heating of dust has an even lower effect than the photodesorption for models run with the commonly-used maximum dust temperature of 70\,K.

In our model, the dominant formation reactions of \hocop\ vary with time. In general, protonation of \cotw\ is more dominant in early time ($<10^5$\,yr), and the gas-phase production with HCO$^+$ and OH becomes more efficient in later times (see Section \ref{sec:intro}). However, the dominant formation routes also vary with physical conditions and it is difficult to conclude which one is more dominant.

Although our models do not include shocks, but we argue that this approach should be sufficient to estimate the \hocop/\cotw\ ratios, especially their upper limit. The \hocop/\cotw\ ratios are determined by the balance among the protonation of \cotw, electron recombination of \hocop, proton transfer from \hocop\ to species with higher proton affinity than \cotw, and the ion-neutral production of \hocop (HCO$^+$ + OH). These reactions occur regardless of shocks. If shocks evaporate \cotw\ significantly, there should be less contribution from the \hocop\ formation through HCO$^+$ and OH compared with the protonation of \cotw, and the \hocop/\cotw\ ratios should be lower, while shocks should increase the fractional abundances of both \hocop\ and \cotw. Therefore, our models without shocks are likely sufficient to obtain upper limits of the \hocop/\cotw\ ratios, but more realistic modeling with shocks will be conducted as future work. It should also be noted that the models without shocks severely underproduce \hocop\ fractional abundances compared with observed peak values, which implies the need for shocks to explain the observed abundances (see Section \ref{sec:disc_origin}).
We include the gas-neutral reaction of HCO$^+$ + OH and many other related reactions in the model. The fact that our model could not reproduce the observed fractional abundances suggests that the formation route through this reaction is not enough to explain our observations.

\section{Discussion}\label{sec:discussion}
\subsection{Origins of \hocop\ emission}\label{sec:disc_origin}
As discussed earlier, the formation routes of \hocop\ do not have to involve the protonation of \cotw. The gas-phase reaction between HCO$^+$ and OH may also contribute to \hocop. \citet{fontani_protonated_2018} argued that \hocop\ must be formed via the reaction above (HCO$^+$ + OH; see Sect. \ref{sec:intro}) in high-mass star-forming cores because \hocop\ fractional abundances derived from the $4_{0,4}-3_{0,3}$ transition are correlated with the fractional abundances of H$^{13}$CO$^+$, while there is no correlation with that of methanol. If \hocop\ is formed via protonation, a large amount of evaporated \cotw\ must be present, which also implies a large amount of methanol in the gas phase. This is because methanol is formed on the ice, and its gas-phase production is extremely inefficient \citep{2007A&A...467.1103G}. 

The CMZ of NGC~253 shows a different trend from the case of these high-mass star-forming cores. We do see a positive spatial correlation between \hocop\ and \methanol\ with low-excitation transitions of both molecules enhanced at the outer CMZ of NGC 253 (Figure \ref{fig:ratio}). Meanwhile, the correlation between \hocop\ and H$^{13}$CO$^+$ is weak because H$^{13}$CO$^+$ is more abundant near the center of the CMZ \citep[Figure \ref{fig:ratio}; see also ][ for abundances of H$^{13}$CO$^+$]{2021ApJ...923...24H}. There is a caveat that the ice composition may be different in high-mass star-forming regions observed by \citet{fontani_protonated_2018} and NGC 253 CMZ, and the presence or lack of correlation may not necessarily imply a difference in formation routes. On the other hand, the presence of a correlation between \methanol\ and \hocop\ and the lack of correlation between \hocop\ and most other species (e.g., CO, HCN, HCO$^+$, etc.) strongly suggest a similar mechanism enhancing abundances of both \methanol\ and \hocop. This mechanism must involve desorption, as methanol is only efficiently formed on ice. Therefore, we argue that \hocop\ in our observations is likely formed from \cotw\ through protonation.

\subsection{Inferred gas-phase \cotw\ fractional abundances}\label{sec:co2frac}
If \hocop\ is produced through the protonation of \cotw\ as we discussed above, \cotw\ must be evaporated from ice into the gas phase because \cotw\ is much more abundant in the ice than in the gas phase (Figure \ref{fig:chem_model}d). Obtaining the gas-phase fractional abundances of \cotw\ could provide essential hints helping us to evaluate the origin of the gas-phase \cotw. From the chemical model, we find that the range of the \hocop/\cotw\ ratio is $\sim 0.001-0.01$. Because the maximum fractional abundance of \hocop\ is $\sim 2\times10^{-9}$ in GMC 1 and $\sim 1\times10^{-9}$ in GMCs 2, 8, and 9, the gas-phase \cotw\ fractional abundances can be \fnum{(1-20)}{-7}  at the outer CMZ, where the \hocop\ intensity peaks.

\subsection{Comparison with ice observations of \cotw}\label{sec:ice}
\begin{figure*}[h]
\centering{
\includegraphics[width=0.49\textwidth]{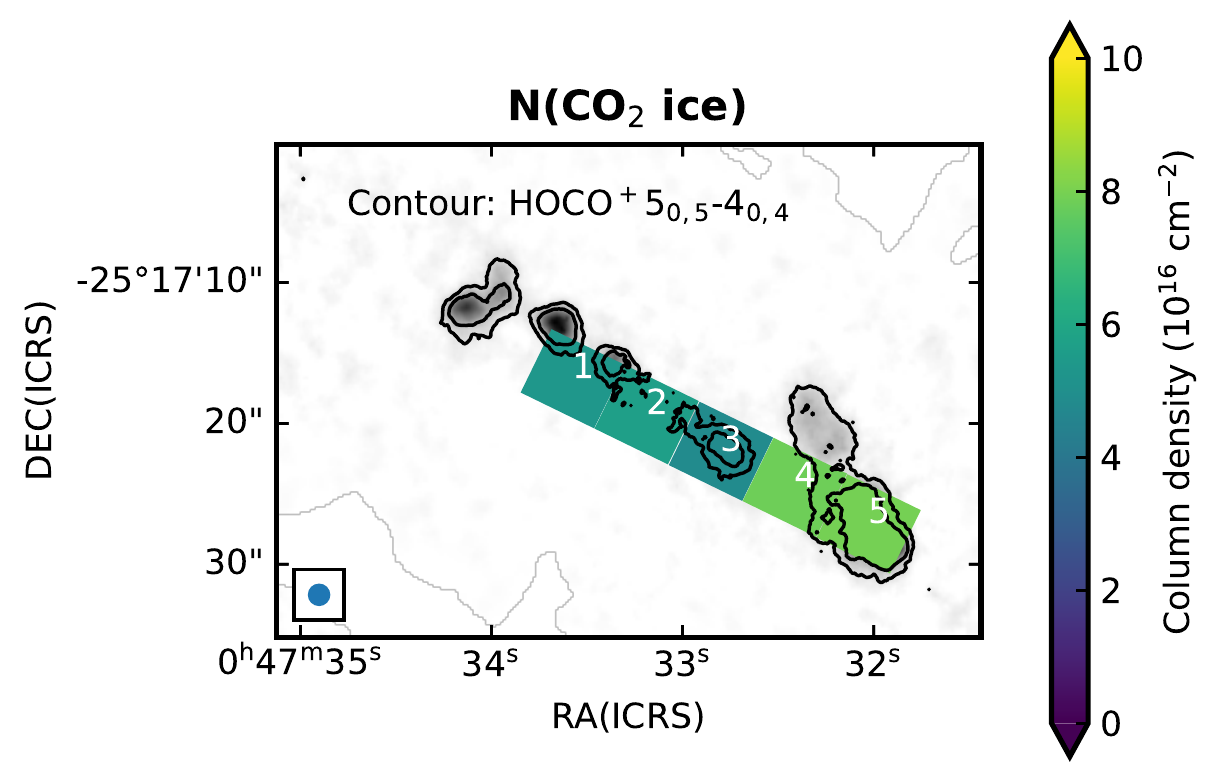}
\includegraphics[width=0.49\textwidth]{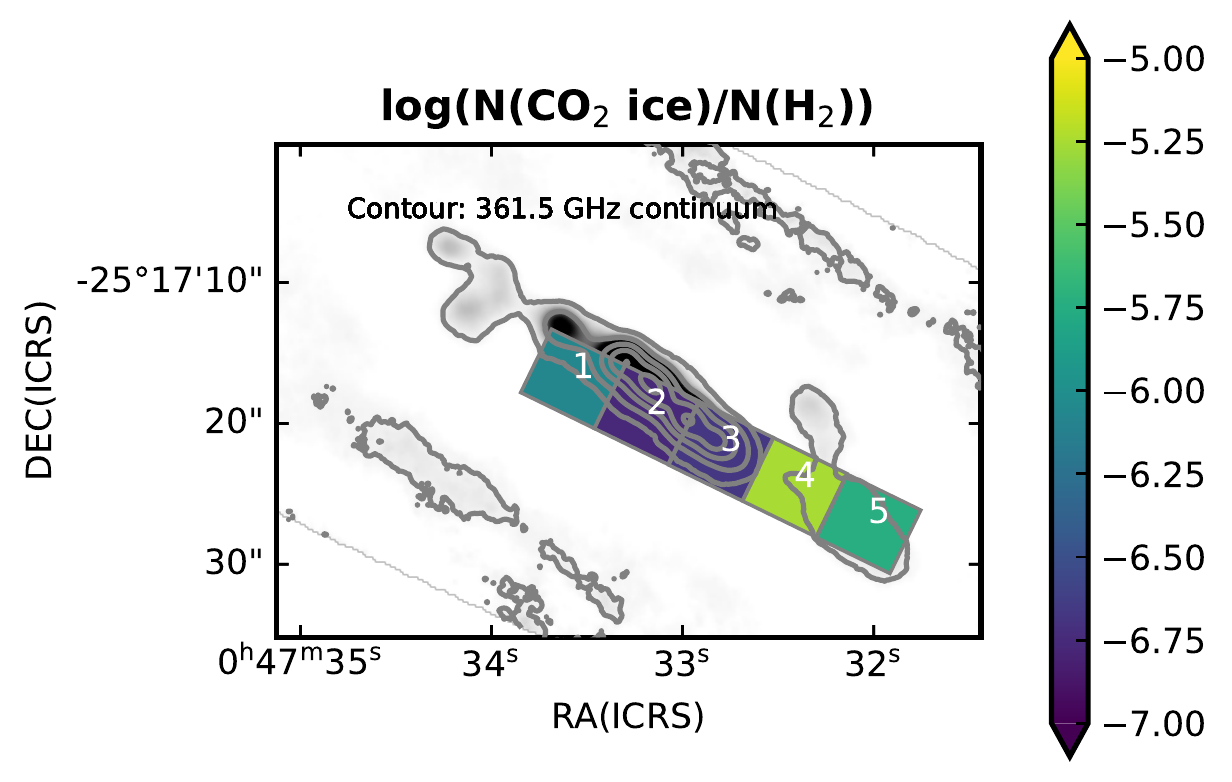}
}
\caption{(Left) Column densities of \cotw\ ice reported by \citet{2015ApJ...807...29Y} are shown for selected slits in color. The integrated intensity image of \hocop\ (5$_{0,5}$-4$_{0,4}$) is shown both in contours and grey scale. (Right) Fractional abundances of \cotw\ ice (i.e., \cotw\ ice column densities divided by the averaged total \htw~column densities inside rectangular regions) are shown in log scale with colored rectangles. The continuum image at 361.5 GHz is shown in contours and grey scale.\label{fig:co2ice}}
\end{figure*}

Here we compare the fractional abundance of \cotw\ gas estimated above with that of \cotw\ ice observations. We utilize the \cotw\ column densities of selected regions of AKARI observations by \citet{2015ApJ...807...29Y}. Although this reference describes these observations in detail, we include the summary of them in Appendix \ref{sec:app_ice}.
These observations used a rectangle slit with a size of $5\arcsec \times 5.8 \arcsec$ (Fig \ref{fig:co2ice} left). Subsequently, we extracted the values of the continuum flux from the same regions to estimate the total \htw\ column density using the method described in Section \ref{sec:coldens}. We then derived the fractional abundances of \cotw\ in the ice phase in these regions shown in Figure \ref{fig:co2ice} (right). We note that the components traced with \cotw\ ice likely come from relatively lower-column-density regions than the ones traced by the dust continuum. Therefore, we have to be aware of the caveat that our estimation of the \cotw\ fractional abundances is rather crude, only accurate for an order of magnitude approximation. 

The \cotw\ ice fractional abundances near the center of NGC 253 (rectangular regions 2 and 3) are lower than in regions 4 and 5 by about an order of magnitude or more. Away from the center, the \cotw\ ice fractional abundances are similar to those of the Milky Way ISM of $\sim 10^{-6}-10^{-5}$ \citep{2003Boonman}, which is also consistent with fractional abundances of \cotw\ ice in our chemical model. Therefore, the \cotw\ ice abundance decreases towards the center of NGC 253, deviating from the Milky Way value. We note that, if we use a higher dust temperature, we would derive lower \htw\ column densities than those we show in Figure \ref{fig:hocop_col}. Subsequently, the fractional abundance estimate would
increase. Although the warm dust does not fully explain the large difference in the derived fractional abundances between the outer part (rectangular regions 4 and 5) and center (rectangular regions 2 and 3) of the CMZ, this uncertainty should be considered in the interpretation of the data.

If the \cotw\ ice fractional abundance is indeed lower in the center than at the outer CMZ, one of the possible factors that may attribute to this suppression is photodissociation, either directly by UV photons or cosmic-ray-induced UV photons. Because of the high star formation rate in the rectangular regions 2 and 3, strong photodissociation is expected \citep{2015ApJ...801...63M}.

Warm dust can also lower the ice abundance of \cotw\ through desorbing \cotw\ ice into the gas phase, or desorbing the precursors of \cotw\ ice. Desorption of many \cotw\ precursors can take place with lower dust temperatures than the desorption of \cotw\ itself. \cotw\ ice is thought to be formed through 
\begin{eqnarray}
    {\rm CO + OH \longrightarrow CO_2 + H}\\
    {\rm CO + O \longrightarrow CO_2}
\end{eqnarray}
\citep{2013A&A...559A..49M} and possibly
\begin{equation}
    {\rm H_2CO + O \longrightarrow CO_2 + H_2}
\end{equation}
\citep{2015A&A...577A...2M}. Some of these reactants have lower binding energies than \cotw\ \citep[$E_b{\rm (CO_2)}\sim 3000\,$K, $E_b{\rm (CO)}\sim 1300\,$K, $E_b{\rm (O)}\sim 1600\,$K;][]{2017MolAs...6...22W,2022arXiv220107512M}\footnote{Note that the binding energy is not the temperature where desorption takes place. Effects of desorption appear when the desorption rate becomes significant enough compared with the accretion rate. There is also an additional complication caused by the ice composition. For example, even if a species has a low binding energy, it may not desorb if it is buried in the ice of another species with a higher binding energy.}. Higher dust temperatures increase the desorption rates of ice species, and species with lower binding energies can desorb with lower dust temperatures \citep[35-50\,K for CO, $\sim 50\,$K for O, $\sim 80\,$K for CO$_2$][]{2022arXiv220107512M}. If a desorption rate of any of these reactants is faster than the reaction rate to form \cotw\, there would be less abundances of \cotw\ ice. Although the mass of warm or hot dust is smaller than that of cold dust, these warm/hot components are likely concentrated in rectangular regions 2 and 3 in Figure \ref{fig:co2ice} where the star formation is active. This can explain the suppression of ice in these regions as well as  low \hocop\ fractional abundances in GMCs 4 and 5 (see Figure \ref{fig:mom0}f). 

Cosmic rays could also contribute to ice desorption. We argued in Section \ref{sec:chem} that cosmic-ray-induced desorption is likely negligible, but its effect could possibly be enhanced in NGC~253. \citet{1993MNRAS.261...83H} estimated that the maximum dust temperature due to the cosmic ray heating of the dust is 70\,K, but this temperature can be different for the case of NGC~253, where some dust is already warm. With the higher maximum dust temperature, the evaporation rate of ice species can be enhanced \citep[e.g., ][]{2020A&A...633A..97K}. This dependence of desorption due to cosmic-ray heating on the dust temperature should be further explored with theoretical studies.

Variation of the \cotw\ ice abundance can be also caused by the initial ice composition. For example, the ice may be rich in atomic or molecular hydrogen. If the ice is abundant in atomic H, frequent hydrogenation reactions can occur, and species such as water, CH$_4$, NH$_3$, and \methanol\ may be abundant. On the other hand, if atomic H is deficient, \cotw\ formation may be a more dominant route of destroying CO ice than \methanol\ formation. Although it is difficult to assess how much the ice compositions differ between Galactic star-forming regions and NGC 253 CMZ, we note that this is another factor that could affect the overall chemistry.

\subsection{Ice and gas-phase chemistry in NGC~253}
We find that the gas-phase fractional abundance of \cotw\ can be $\sim (1-20)\times 10^{-7}$ in GMCs 1, 2, 8, and 9 (Section \ref{sec:co2frac}), and the ice fractional abundance is $\sim 2\times 10^{-6}$ around GMC 1 (rectangular region 5 in Figure \ref{fig:co2ice}). This means that there is a process sublimating a large fraction of \cotw\ ice into the gas phase in these GMCs. 

One possible \cotw\ sublimation mechanism is a hydrodynamical shock. Chemical models have shown that shocks can sputter off the ice because some gas particles have enough kinetic energy to desorb ice  \citep{2008A&A...482..549J,2011ApJ...740L...3V,2012MNRAS.421.2786F}. We also note that shock sublimation occurs even when the dust temperature is low, because the energy is provided by gas. When the shock velocity is high $>20\,$km\,s$^{-1}$, ice sputtering is efficient enough to desorb a large fraction of ice \citep{2015A&A...584A.102H}. Ubiquitous methanol emission is likely attributed to shock sublimation in the Milky Way Galactic Center \citep{2009ApJ...692...47M}, and this type of ice sublimation likely occurs also in NGC 253 with frequent shocks \citep{2015ApJ...801...63M}.  The locations with enhanced \hocop\ abundances are considered as intersections of different orbits.  As shown in Figure \ref{fig:orbit}, bar orbits ($x_1$ orbits) and inner orbits ($x_2$ orbits) intersect at the northeast and southwest parts of the CMZ, near GMCs 1, 8, 9, and 10. Shocks at GMCs 1, 2, 7, 8, 9, and 10 have been suggested by the detection of Class~I methanol masers \citep{2022arXiv220503281H}, and some other molecular tracers of shocks (Huang et al. in preparation). Because these regions are located at intersections of different orbits \citep[bar and nuclear ring; see ][ for the dynamical modeling]{2000PASJ...52..785S,2001ApJ...549..896D,2022arXiv220604700L}, shocks due to cloud collisions are not surprising.

If the shock scenario is correct, ice sputtering of other species in addition to \cotw\ should be taking place. Methanol enhancement shown in Section \ref{sec:integint} is one example supporting this scenario. Water is another species abundant in ice, and ice sputtering should increase its gas-phase abundance. From the \water/\cotw\ ratio of $\sim 7$ in ice \citep{2015ApJ...807...29Y} and our estimated gas-phase fractional abundance of \cotw\ of $(1-20)\times 10^{-7}$, the gas-phase water abundances in shocked regions must be $\sim (0.7-14)\times 10^{-6}$. This estimate assumes that the same fraction of \cotw\ and \water\ ice is sublimated, and does not consider the higher desorption energy of water compared with \cotw. It also assumes that gas-phase reactions subsequent to sputtering do not change the \cotw\/\water\ ratio. \citet{2017ApJ...846....5L} derived fractional abundances of gas-phase water to be $\sim 10^{-7}$ from Herschel HIFI/PACS/SPIRE data using multiple transitions of water in analysis with all data convolved to $40\arcsec$ at the center of NGC 253. This value may be locally higher in shocked regions. However, it is impossible to confirm it without spatially resolving shocked regions and nuclear starburst regions with a higher-angular-resolution ($<10\arcsec$) observation.

Another possible \cotw\ sublimation mechanism is thermal desorption. We do not know the distribution of the dust temperature due to the lack of high angular resolution data at the wavelengths of the peak black body radiation intensity ($\sim 10-100\,\mu$m for 30-300\,K). Yet, we do know that the most active star formation takes place around GMCs 3-5 from radio recombination line or 3mm continuum data \citep{2015MNRAS.450L..80B, 2021ApJ...923...24H}. If the hot/warm dust components are concentrated in GMCs 3-5, it is unlikely that the thermal desorption already takes place at GMCs 1 and 9. Therefore, thermal desorption unlikely contributes to the sublimation of \cotw\ in GMCs 1, 2, 8, and 9. However, it is quite possible that thermal desorption occurs around GMC 3-5 due to the high dust temperature.

Another mechanism which can sublimate \cotw\ is cosmic-ray-induced desorption as described in Section \ref{sec:ice}. We argued that this desorption mechanism is unlikely unless the dust is warm so that the maximum dust temperature achieved from cosmic-ray heating of dust becomes significantly larger than 70\,K. Without high star formation rates in GMCs 1 and 9, it is unlikely that the dust is already warm.

For the reasons above, we conclude that a shock is the most likely scenario driving the \cotw\ evaporation. Yet, to better constrain the ice fractional abundances, high angular resolution observations at infrared wavelengths, e.g., from JWST, are crucial.

\section{Summary}\label{sec:summary}
In this paper, we analyzed the abundances of \hocop,  the protonated form of \cotw, in the central molecular zone of the starburst galaxy NGC~253, and discussed its relationship with the gas-and ice-phase \cotw. Below is the summary of our findings.
\begin{itemize}
    \item The distribution of \hocop\ shows clear enhancements at locations of $x_1$ and $x_2$ orbital intersections where shocks are expected. This distribution is similar to that of methanol but is different from that of \httcop. There are two formation routes of \hocop; one is the ion-neutral reaction HCO$^+$ + OH and the other is the protonation of \cotw. If the former route is dominant, the \hocop\ distribution should be similar to that of HCO$^+$, while the latter route should cause similarity with the \methanol\ distribution.  Therefore, \hocop\ is likely produced through the protonation of \cotw.
    \item We derive \hocop\ column densities across the CMZ using CASSIS, from which we also obtain its fractional abundances using the total H$_2$ column densities estimated from the dust emission. We find \hocop\ fractional abundances as high as $\sim 2\times 10^{-9}$, which is similar to those observed in the Galactic center, but orders of magnitude higher than those reported in Galactic spiral-arm molecular clouds. 
    \item From the results of chemical modeling and values of cosmic-ray ionization rates derived from previous ALCHEMI works, we estimate that the gaseous \hocop/\cotw\ ratio is likely $10^{-3} - 10^{-2}$. This ratio suggests that the gas-phase \cotw\ fractional abundances are \fnum{(1-20)}{-7} at peaks of \hocop\ emission.
    \item We also estimate fractional abundances of \cotw\ ice from their column densities in the literature. The ice fractional abundance at the \hocop\ peak is similar to the value in the Galactic interstellar medium ($10^{-6}-10^{-5}$), but is lower ($\sim (1-3)\times 10^{-7}$) near the NGC~253 galactic center.   
    \item The increased gaseous and ice fractional abundances of \cotw\ at the outer CMZ of NGC~253 imply that a large fraction of ice is sublimated. Because of the association of these locations with evidence of shocks, we propose that this efficient sublimation is attributed to shock-induced sputtering. 
\end{itemize}
High spatial resolution observations of molecular emission in external galaxies, such as those performed by the ALCHEMI survey toward the central regions of the starburst galaxy NGC~253 with ALMA, have greatly improved our understanding of gas-phase abundances and will continue to do so. Now, complementary observations of ice at high angular resolutions with the JWST are required to obtain a complete picture of the chemical processes in starburst galaxies.

\begin{acknowledgements}
\sloppypar{We thank the anonymous referee for constructive comments. N.H. thanks Hideko Nomura for the helpful discussion on \cotw\ transitions in the infrared wavelength, and Kotomi Taniguchi for the initial help using CASSIS. This paper makes use of the following ALMA data: ADS/JAO.ALMA\#2017.1.00161.L, ADS/JAO.ALMA\#2018.1.00162.S, ADS/JAO.ALMA\#2018.1.01321.S.} ALMA is a partnership of ESO (representing its member states), NSF (USA) and NINS (Japan), together with NRC (Canada), MOST and ASIAA (Taiwan), and KASI (Republic of Korea), in cooperation with the Republic of Chile. The National Radio Astronomy Observatory is a facility of the National Science Foundation operated under cooperative agreement by Associated Universities, Inc. The Joint ALMA Observatory is operated by ESO, AUI/NRAO and NAOJ. Data analysis was in part carried out on the Multi-wavelength Data Analysis System operated by the Astronomy Data Center (ADC), National Astronomical Observatory of Japan. N.H. acknowledges support from JSPS KAKENHI Grant Number JP21K03634. V.M.R. has received support from the Comunidad de Madrid through the Atracci\'on de Talento Investigador Modalidad 1 (Doctores con experiencia) Grant (COOL:Cosmic Origins of Life; 2019-T1/TIC-5379), and the Ayuda RYC2020-029387-I funded by MCIN/AEI /10.13039/501100011033. L.C. has received partial support from the Spanish State Research Agency (AEI; project number PID2019-105552RB-C41). P.H. is a member of and received financial support for this research from the International Max Planck Research School (IMPRS) for Astronomy and Astrophysics at the Universities of Bonn and Cologne. 
K.S. is supported by the grant MOST 109-2112-M-001-020 from the Ministry of Science and Technology, Taiwan. This research has made use of the NASA/IPAC Extragalactic Database (NED), which is funded by the National Aeronautics and Space Administration and operated by the California Institute of Technology.
\end{acknowledgements}

\facilities{ALMA}
\software{Astropy \citep{astropy:2013, astropy:2018}, CASA \citep{casa}, CASSIS \citep{2015sf2a.conf..313V}, Nautilus \citep{2016MNRAS.459.3756R}}
\appendix
\section{CASSIS fitting parameters}\label{sec:app_cassis}
CASSIS constrains parameters such as column density, line velocity, line width (full width at half maximum; FWHM), excitation temperatures, and source sizes by fitting the spectra. We used the MCMC algorithm provided by CASSIS to derive these parameters. We use spectra from transitions shown in the upper half of Table \ref{tab:spec}. When using the MCMC method, users provide acceptable ranges of these parameters as well as the initial guesses. We used the column density range of [$10^{12}$,$10^{16}$] cm$^{-2}$ with the initial guess of \enum{14} cm$^{-2}$, the excitation temperature range [5,50] K with the initial guess of 6\,K, the velocity range [$v_{\rm CO} -30$,$v_{\rm CO}+30.$]\,\kms\ where $v_{\rm CO}$ is the velocity obtained from the moment 1 image of CO(1-0), which is also used for the initial guess. The range of FWHM we used is [$\sigma_V$ -20.,90.]\,\kms\ with the initial guess of $\sigma_V$ where $\sigma_V$ is the velocity dispersion of the CO(1-0) image\footnote{The relationship between the FWHM and the standard deviation of a Gaussian distribution is usually described as $FWHM \sim 2.355 \sigma_V$. In our case, the line width of \hocop\ is much smaller than that of CO(1-0), and the initial guess of $\sigma_V$ for the FWHM of \hocop\ is still a reasonable one.}. We note that the line width of \hocop\ is significantly different from that of CO(1-0), and this range of FWHM is simply determined by running CASSIS multiple times and checking the range to produce a reasonable fit. 

\begin{figure*}[h]
\centering{
\includegraphics[width=0.49\textwidth]{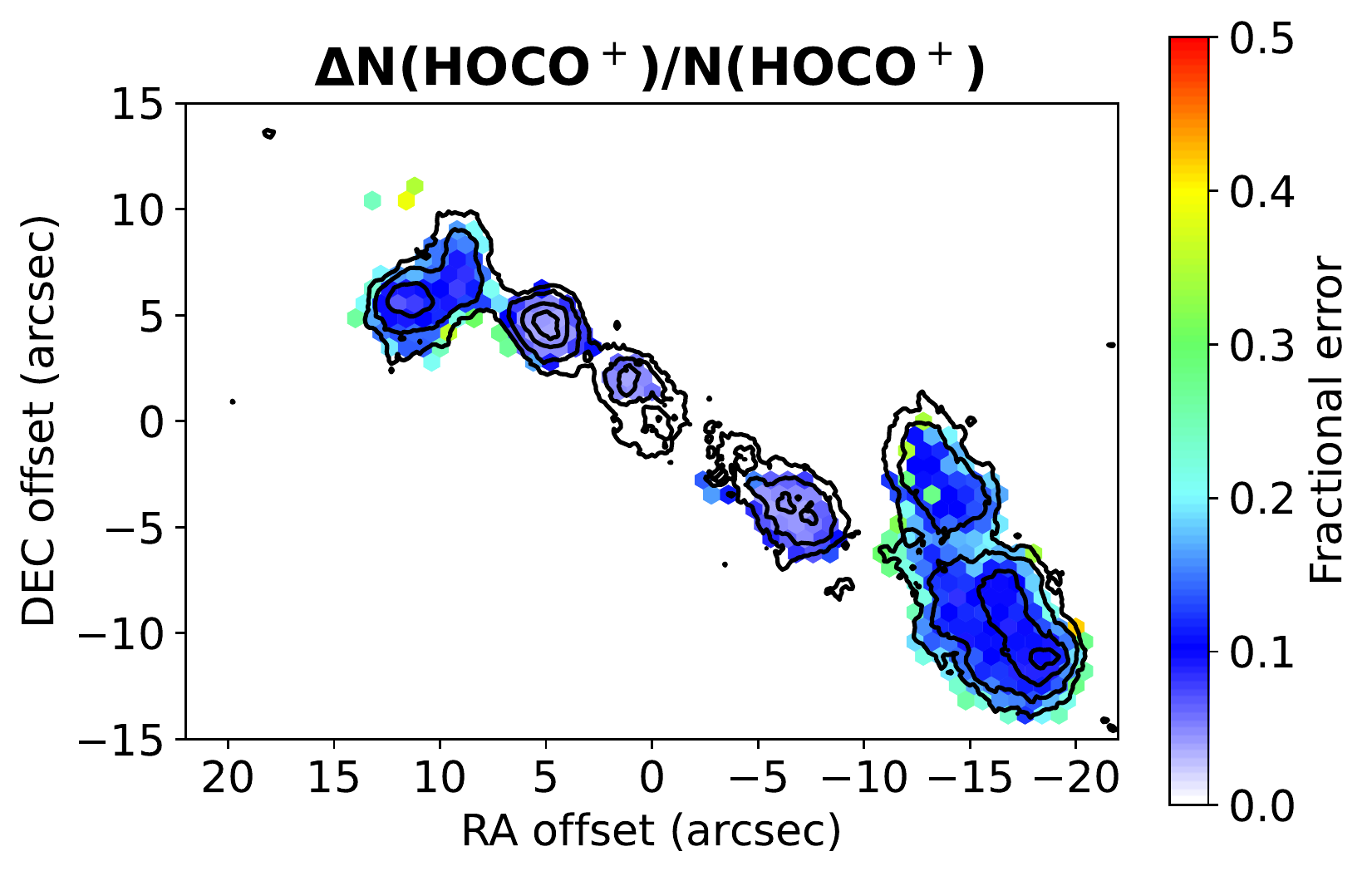}
\includegraphics[width=0.49\textwidth]{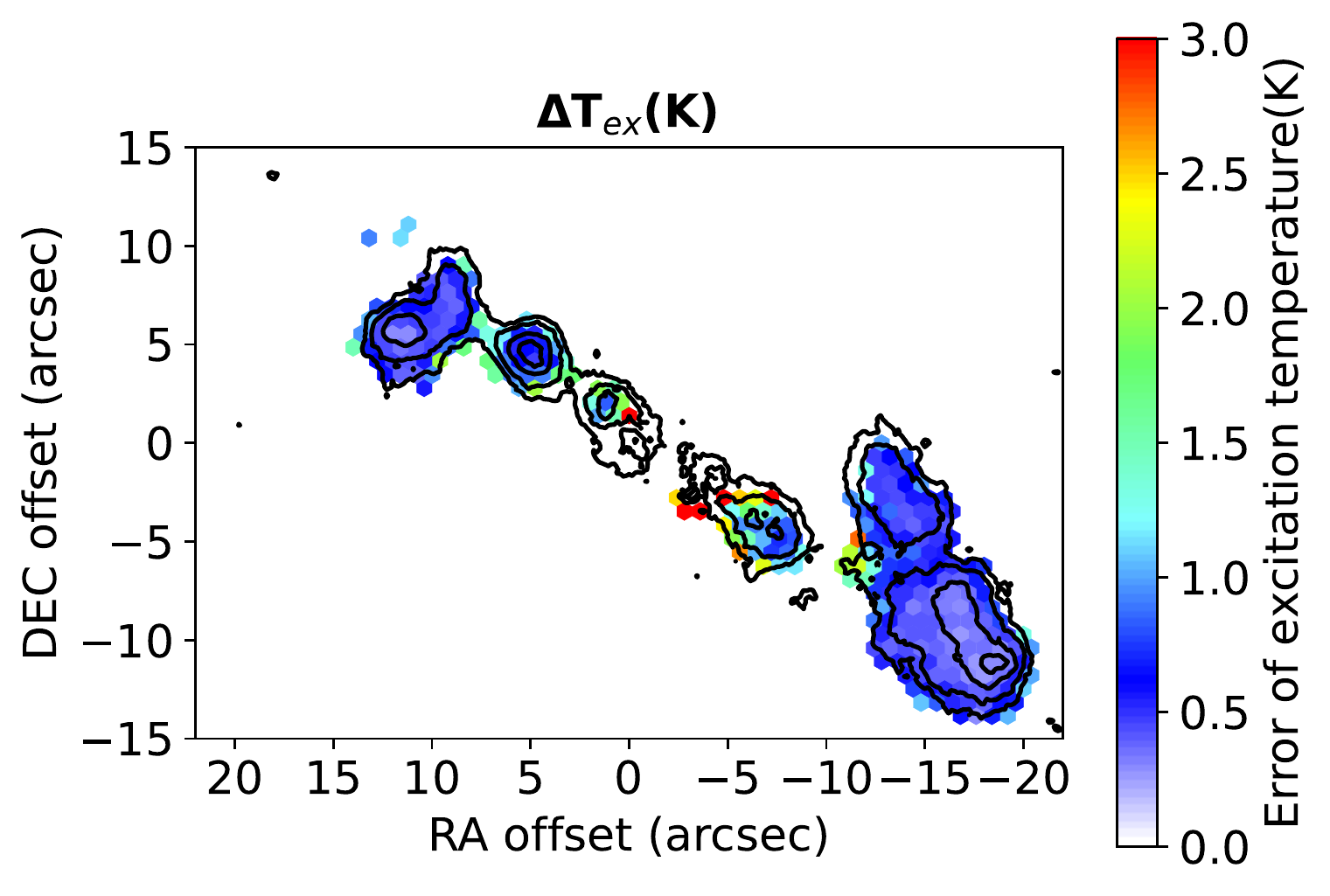}}
\caption{(Left) Ratios of errors in the derivation of \hocop\ column densities over the derived \hocop\ column densities at corresponding hexagonal pixels. (Right) Errors of excitation temperatures. \label{fig:error_cassis}}
\end{figure*}

Uncertainties from the CASSIS fit are reasonably small for most cases. Figure \ref{fig:error_cassis}(left) shows that the errors of the column densities  are around 10\,\% for most cases, and $\sim 20-30\,$\% for a small fraction of pixels with low signal-to-noise ratios. Uncertainties in excitation temperatures are within 0.5\,K for most pixels, with a maximum of 3\,K.

There are also other sources of uncertainties, in addition to the spectral fitting. For example, \hocop\ column densities derived with spectroscopic constants from Jet Propulsion Laboratory (JPL; https://spec.jpl.nasa.gov/) would yield up to a factor of 2 larger values compared with ones from CDMS. Although this is a large factor, a use of different spectroscopic constants changes the results uniformly within the field of view. There are also observational uncertainties of up to 15\% \citep{2021A&A...656A..46M}. These uncertainties will not change our main conclusions.

The \hocop\ transitions are optically thin even where the column densities are high. CASSIS does not provide the optical depths, but we also ran MADCUBA 
\citep{2019A&A...631A.159M}, a similar spectral fitting software to obtain optical depths. The optical depths are $<0.1$ for positions that we checked, which have high intensities of \hocop.

The modeled intensities with CASSIS for pixels that are closest to each GMC position are shown in Table \ref{tab:model_int} if \hocop\ transition for a GMC is detected.


\begin{deluxetable*}{ccccccccc} 
\tablecolumns{9} 
\tablewidth{0pc} 
\tablecaption{Modeled integrated intensities\label{tab:model_int}}
    \tablehead{\colhead{GMC} &\colhead{$x_{\rm offset}$} &\colhead{$y_{\rm offset}$} &\multicolumn{6}{c}{$\int I dv$}\\
\colhead{} &\colhead{($\arcsec$)} &\colhead{($\arcsec$)} &\multicolumn{6}{c}{(K\,km\,\ps)}}
\startdata 
&&&($4_{0,4}-3_{0,3}$)&($5_{0,5}-4_{0,4}$)&($6_{0,6}-5_{0,5}$)&($7_{0,7}-6_{0,6}$)&($8_{0,8}-7_{0,7}$) &($12_{0,12}-11_{0,11}$)\\
\cline{4-9}\\
1 &-17.3, &-10.4 &12.14 &9.00 &5.25 &2.47 &0.94 &0.00 \\
2 &-13.6, &-2.8 &6.40 &4.93 &2.92 &1.39 &0.55 &0.00 \\
3 &-6.5,& -4.2 &6.39 &7.99 &8.96 &9.06 &8.36 &3.02 \\
6 &0.4, &2.1 &4.41 &6.05 &7.42 &8.37 &8.81 &6.28 \\
7 &4.8,& 4.2 &11.12 &12.40 &11.84 &9.95 &7.48 &0.88 \\
8 &10.0, &6.2 &7.30 &6.05 &4.03 &2.22 &1.01 &0.01 \\
9 &12.0,& 5.6 &13.08 &10.73 &7.11 &3.85 &1.73 &0.01 \\
\enddata 
\tablecomments{Integrated intensities of \hocop\ transitions produced by CASSIS fitting for hexagonal pixels closest to GMCs. Results are shown only for GMCs with \hocop\ detection. \label{tab:gmc_intens}} 
\end{deluxetable*} 

\section{Ice data from AKARI}\label{sec:app_ice}
Here we summarize analyses by \citet{2015ApJ...807...29Y}, and present spectra for regions used in our analysis. The observed wavelength range is about $2.5-5.0\,\mu$m. Within this range, CO$_2$ ice, H$_2$O ice, Br$\alpha$, and PAH $3.3\,\mu$m features were detected in addition to the continuum. \cotw\ ice features at $4.27\,\mu$m are fit using the data range of $4.1-4.4\,\mu$m as there is only this narrow ice feature in this wavelength range. Because the ice composition changes the spectral shape, multiple ice compositions were tested to best fit the spectra. Consequently, the ice composition of \water:\methanol:\cotw=$9:1:2$ was used in the final analysis.

\begin{figure*}[h]
\centering{
\includegraphics[width=0.99\textwidth]{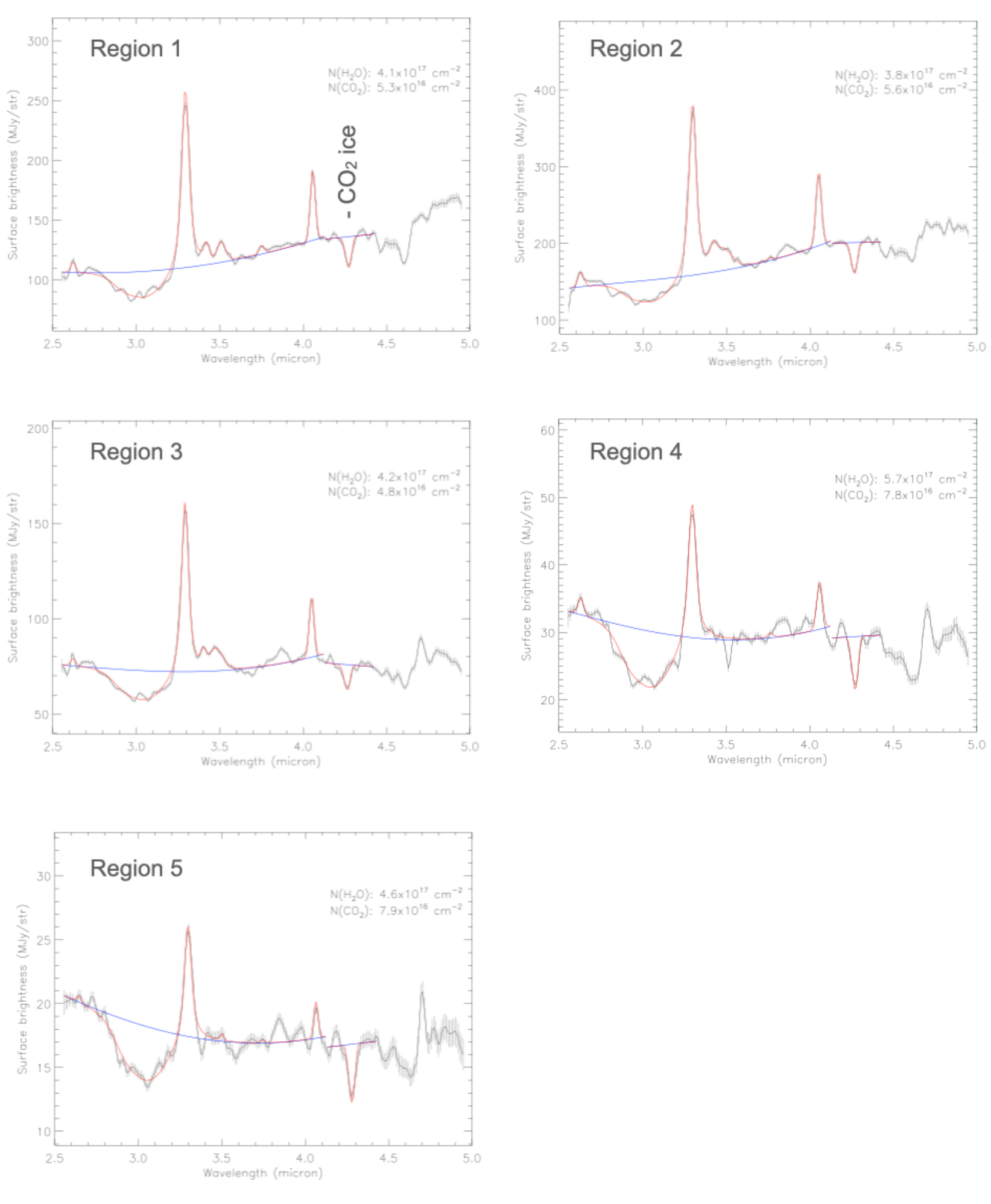}
}
\caption{AKARI spectra for regions used in our analysis. As in Figure 1 of \citet{2015ApJ...807...29Y}, the blue curve represents the best-fit continuum emission while the red curve shows the overall spectral profiles. \label{fig:ice_spec}}
\end{figure*}

Figure \ref{fig:ice_spec} shows spectra used to derive \cotw\ ice column densities shown in Figure \ref{fig:co2ice}. Note that our Regions 1-5 correspond to ID 47-51 in Table 3 of \citet{2015ApJ...807...29Y}. 

\section{GMC positions}\label{sec:app_gmc}
As already noted in \citet{2022arXiv220503281H} and Behrens et al. (submitted to ApJ), the GMC nomenclature was adopted from \citet{leroy_alma_2015}, but with modified positions. How these modifications are made is explained in Behrens et al. (submitted to ApJ).
These positions are shown in Table \ref{tab:gmc}.

\begin{deluxetable*}{ccc} 
\tablecolumns{3} 
\tablewidth{0pc} 
\tablecaption{GMC positions\label{tab:gmc}}
\tablehead{\colhead{ID} &\colhead{RA (ICRS)} &\colhead{DEC (ICRS)}\\
\colhead{} &\colhead{$00^h47^m-^s$} &\colhead{$-25^\circ17'-''$}}
\startdata 
1 & 32.02 & 28.2 \\
2 & 32.28 & 20.2 \\
3 & 32.81 & 21.6 \\
4 & 32.97 & 20.0 \\
5 & 33.21 & 17.4 \\
6 & 33.33 & 15.8 \\
7 & 33.64 & 13.3 \\
8 & 34.02 & 11.4 \\
9 & 34.17 & 12.3 \\
10 & 34.24 & 7.8 \\
\enddata 
\tablecomments{Modified coordinates of GMC positions in \citet{leroy_alma_2015} provided by A. K. Leroy (private communication).} 
\end{deluxetable*} 

\section{The archival CO(2-1) image}
The large-scale CO(2-1) image of NGC 253 shown in Figure \ref{fig:orbit} was taken from the ALMA archive (project code \#2018.1.01321.S). These data use the configuration consisting on the 7-m array complemented by the total power antenna. Pipeline-reduced image cubes (QA2 products) for the 7-m array and total power single-dish data were combined with the CASA command {\tt feather}. This image was shown to indicate the rough positions of $x_1$ and $x_2$ orbits only. We expect that the PI team will present the data with better imaging quality and scientific analysis.


\input{ALCHEMI_HOCOp.bbl}


\end{document}

%% file: ALCHEMICollabAuthList.tex
\author[0000-0002-6824-6627]{Nanase Harada}
\affiliation{National Astronomical Observatory of Japan, 2-21-1 Osawa, Mitaka, Tokyo 181-8588, Japan}
\affiliation{Department of Astronomy, School of Science, The Graduate University for Advanced Studies (SOKENDAI), 2-21-1 Osawa, Mitaka, Tokyo, 181-1855 Japan}


\author[0000-0001-9281-2919]{Sergio Mart\'in}
\affiliation{European Southern Observatory, Alonso de C\'ordova, 3107, Vitacura, Santiago 763-0355, Chile}
\affiliation{Joint ALMA Observatory, Alonso de C\'ordova, 3107, Vitacura, Santiago 763-0355, Chile}
%
\author[0000-0003-1183-9293]{Jeffrey G.~Mangum}
\affiliation{National Radio Astronomy Observatory, 520 Edgemont Road,
  Charlottesville, VA  22903-2475, USA}

\author[0000-0001-5187-2288]{Kazushi Sakamoto}
\affiliation{Institute of Astronomy and Astrophysics, Academia Sinica, 11F of AS/NTU
Astronomy-Mathematics Building, No.1, Sec. 4, Roosevelt Rd, Taipei 10617, Taiwan}

\author[0000-0002-9931-1313]{Sebastien Muller}
\affiliation{Department of Space, Earth and Environment, Chalmers University of Technology, Onsala Space Observatory, SE-439 92 Onsala, Sweden}

\author[0000-0002-2887-5859]{V\'ictor M. Rivilla}
\affiliation{Centro de Astrobiolog\'ia (CSIC-INTA), Ctra. de Ajalvir Km. 4, Torrej\'on de Ardoz, 28850 Madrid, Spain}

\author[0000-0002-7495-4005]{Christian Henkel}
\affiliation{Max-Planck-Institut f\"ur Radioastronomie, Auf dem H\"ugel
  69, 53121 Bonn, Germany}
\affiliation{Astronomy Department, Faculty of Science, King Abdulaziz
  University, P.~O.~Box 80203, Jeddah 21589, Saudi Arabia}
  \affiliation{Xinjinag Astronomical Observatory, Chinese Academy of Sciences,
   830011 Urumqi, PR China}

\author[0000-0001-9436-9471]{David S.~Meier}
\affiliation{New Mexico Institute of Mining and Technology, 801 Leroy Place, Socorro, NM 87801, USA}
\affiliation{National Radio Astronomy Observatory, PO Box O, 1003 Lopezville Road, Socorro, NM 87801, USA}

\author[0000-0001-8064-6394]{Laura Colzi}
\affiliation{Centro de Astrobiolog\'ia (CSIC-INTA), Ctra. de Ajalvir Km. 4, Torrej\'on de Ardoz, 28850 Madrid, Spain}

\author[0000-0002-6385-8093]{Mitsuyoshi Yamagishi}
\affiliation{Institute of Astronomy, Graduate School of Science,
The University of Tokyo, 2-21-1 Osawa, Mitaka, Tokyo 181-0015, Japan}

\author[0000-0001-8153-1986]{Kunihiko Tanaka}
\affil{Department of Physics, Faculty of Science and Technology, Keio University, 3-14-1 Hiyoshi, Yokohama, Kanagawa 223--8522 Japan}

\author[0000-0002-6939-0372]{Kouichiro Nakanishi}
\affiliation{National Astronomical Observatory of Japan, 2-21-1 Osawa, Mitaka, Tokyo 181-8588, Japan}
\affiliation{Department of Astronomy, School of Science, The Graduate University for Advanced Studies (SOKENDAI), 2-21-1 Osawa, Mitaka, Tokyo, 181-1855 Japan}

\author[0000-0002-7758-8717]{Rub\'en~Herrero-Illana}
\affiliation{European Southern Observatory, Alonso de C\'ordova 3107, Vitacura, Casilla 19001, Santiago de Chile, Chile}
\affiliation{Institute of Space Sciences (ICE, CSIC), Campus UAB, Carrer de Magrans, E-08193 Barcelona, Spain}

\author[0000-0002-1413-1963]{Yuki Yoshimura}
\affiliation{Institute of Astronomy, Graduate School of Science,
The University of Tokyo, 2-21-1 Osawa, Mitaka, Tokyo 181-0015, Japan}

\author[0000-0003-3537-4849]{P. K. Humire}
\affiliation{Max-Planck-Institut f\"ur Radioastronomie, Auf dem H\"ugel 69, 53121 Bonn, Germany}

\author[0000-0002-1316-1343]{Rebeca Aladro}
\affiliation{Max-Planck-Institut f\"ur Radioastronomie, Auf dem H\"ugel 69, 53121 Bonn, Germany}

\author[0000-0001-5434-5942]{Paul P.~van der Werf}
\affiliation{Leiden Observatory, Leiden University, PO Box 9513, NL-2300 RA Leiden, The Netherlands}

\author[0000-0001-6527-6954]{K. L. Emig}
\altaffiliation{Jansky Fellow of the National Radio Astronomy Observatory}
\affiliation{National Radio Astronomy Observatory, 520 Edgemont Road, Charlottesville, VA 22903-2475, USA}